\documentclass[12pt]{article}

\usepackage{amsmath,amsthm,amssymb,mathrsfs}
\usepackage[linesnumbered,ruled]{algorithm2e}
\usepackage{bm}
\usepackage[dvipsnames]{xcolor}
\usepackage{enumerate}
\usepackage{fullpage}
\usepackage{graphicx}
\usepackage[colorlinks=true]{hyperref}
\usepackage{multirow}
\usepackage{natbib}
\usepackage{rotating}
\usepackage{subfig}
\usepackage{url}

\hypersetup{
    colorlinks=true,
    linkcolor=Maroon,   
    urlcolor=Green,
    citecolor=NavyBlue,
}

\SetKwInOut{Parameter}{Parameters}

\newtheorem{theorem}{Theorem}
\newtheorem{lemma}{Lemma}

\theoremstyle{definition}
\newtheorem{condition}{Condition}

\newtheorem{example}{Example}
\newtheorem{remark}{Remark}

\def\spacingset#1{\renewcommand{\baselinestretch}%
{#1}\small\normalsize} \spacingset{1}

\newcommand{\argmin}{\operatorname*{\arg\min}}

\newcommand{\E}{\operatorname{{E}}} 
\newcommand{\I}{\textnormal{I}} 
\newcommand{\Var}{\operatorname{Var}}

\newcommand{\pa}{\textnormal{\textsc{pa}}}

\newcommand{\vx}{{V}}
\newcommand{\ARG}{\textsc{arg}}

\newcommand{\sign}{\operatorname{sign}} 
\newcommand{\diag}{\operatorname{Diag}} 
\newcommand{\topdep}{{d}} 

\newcommand{\blind}{0}

\newcommand{\Title}{\Large\bf Nonlinear causal discovery with confounders}

\begin{document}

\if0\blind
{
\title{\Large\bf \Title 
\thanks{Corresponding author: C.~Li (\url{li000007@umn.edu}). $^1$School of Statistics, $^2$Division of Biostatistics, University of Minnesota, Minneapolis, MN 55455. 
The research is supported in part by NSF grant DMS-1952539, NIH grants R01GM113250, R01GM126002, R01AG065636, R01AG074858, R01AG069895, U01AG073079. 
The authors would like to thank the editor, the associate editor, and the anonymous referee for their helpful comments and suggestions.
}}
\author{Chunlin Li$^1$ \and Xiaotong Shen$^1$ \and Wei Pan$^2$}
\date{}
\maketitle
} \fi
\if1\blind
{
\title{\Large\bf \Title }
\author{}
\date{}
\maketitle
} \fi

\begin{abstract}\footnotesize
This article introduces a causal discovery method to learn nonlinear relationships in a directed acyclic graph with correlated Gaussian errors due to confounding. First, we derive model identifiability under the sublinear growth assumption. Then, we propose a novel method, named the Deconfounded Functional Structure Estimation (DeFuSE), consisting of a deconfounding adjustment to remove the confounding effects and a sequential procedure to estimate the causal order of variables. We implement DeFuSE via feedforward neural networks for scalable computation. Moreover, we establish the consistency of DeFuSE under an assumption called the {strong causal minimality}. In simulations, DeFuSE compares favorably against state-of-the-art competitors that ignore confounding or nonlinearity. Finally, we demonstrate the utility and effectiveness of the proposed approach with an application to gene regulatory network analysis. The Python implementation is available at \url{https://github.com/chunlinli/defuse}.
\end{abstract}


\spacingset{1.5} 

\section{Introduction}

Causal relationships are fundamental to understanding the mechanisms of complex systems and the consequences of actions in natural and social sciences. 
Causal discovery, namely to learn a directed acyclic graph (DAG) representing causal relationships, arises in many applications.
In gene network analysis, scientists explore gene-to-gene regulatory relationships 
to unravel the genetic underpinnings of a disease \citep{sachs2005causal}.
In such a situation, latent confounders such as environmental or lifestyle factors
could introduce spurious associations or mask causal relationships in observed gene expression levels, making causal discovery more challenging.
Currently, causal discovery from observational data is an important research topic as randomized experiments are often unethical, expensive, or infeasible. 
In this paper, we concentrate on the discovery of causal relationships in the presence of latent confounders. 

Linear causal discovery without confounders has been extensively studied 
\citep{spirtes2000causation, chickering2002optimal,tsamardinos2006max, shimizu2006linear, de2006scoring, jaakkola2010learning,de2011efficient, gu2019penalized, zheng2018dags,yuan2019constrained,li2020likelihood}. 
However, in practice, many causal relations are nonlinear, 
raising concerns about using a linear model \citep{voorman2014graph}. 
For nonlinear causal models without confounders, three major approaches include 
(1) nonlinear independent component analysis \citep{monti2020causal,zhang2009identifiability}, 
(2) combinatorial search for the causal order \citep{mooij2009regression,buhlmann2014cam}, and 
(3) continuous constrained optimization for causal structure learning \citep{zheng2020learning}. 
The first estimates the functional relations through the mutual independence of errors. 
The second determines the causal order based on a certain criterion. 
For example, the causal additive model (CAM) \citep{buhlmann2014cam} assumes the nonlinear functions are of additive form and estimates the causal order that maximizes the likelihood.
The third approach directly optimizes an objective function subject to a smooth constraint characterizing acyclicity.
The most representative example is NOTEARS \citep{zheng2020learning}. 
The reader may consult \citet{peters2017elements} and \citet{glymour2019review} for excellent surveys of nonlinear causal discovery.

In the presence of latent confounders, several methods are available for linear causal discovery.
As extensions of the PC algorithm, FCI \citep{spirtes2000causation} and its variant RFCI \citep{colombo2012learning} address latent confounders by producing a partial ancestral graph (PAG) instead of a completed partially DAG (CPDAG).
Another approach \citep{frot2019robust,shah2020right} assumes the confounding is pervasive \citep{chandrasekaran2012latent,wang2019blessings} and recovers the CPDAG in two steps. 
For example, LRpS-GES \citep{frot2019robust} uses the low-rank plus sparse estimator \citep{chandrasekaran2012latent} to remove confounding, followed by the GES algorithm \citep{chickering2002optimal} to perform causal structure estimation. 
Besides, the instrumental variable estimation is a well-known approach but requires the availability of valid instruments \citep{chen2018two,li2021inference}. 

Despite the foregoing progress, nonlinear causal discovery with confounders remains largely unexplored.
In a bivariate case, the work of \citet{janzing2009identifying} estimates the confounding effect by minimizing the $L_2$-distance between data points and a curve evaluated at the estimated values of the confounder. 
For a multivariate case, it remains unclear whether nonlinearity can help causal discovery with confounding, 
although third-order differentiability suffices for the identifiability of nonlinear causal discovery without confounders \citep{peters2014causal}. 
Moreover, major computational and theoretical challenges arise when we confront the curse of dimensionality in learning a nonparametric DAG.
During the review process, a preprint by \citet{agrawal2021decamfounder} proposes a two-step procedure for nonlinear causal discovery in the presence of pervasive confounders. However, for consistent estimation, their method requires that the sample size grows slower than the quadratic graph size, $n\ll p^2$, which may be restrictive, especially for nonparametric estimation.

This paper contributes to the following areas. 
First, we derive a new condition, called the \emph{sublinear growth assumption}, for model identifiability in the presence of latent confounders. 
Second, we propose a novel approach for causal discovery, called the Deconfounded Functional Structure Estimation (DeFuSE), comprising a deconfounding adjustment and an iterative procedure to reconstruct the topological order of the variables. 
Third, we implement DeFuSE through feedforward neural networks without assuming additive functional relationships while allowing efficient computation for a reasonable graph size $p$, say $p=100$. 
This is in contrast to traditional nonparametric methods that suffer from inefficiency in high dimensions, such as tensor-product B-splines \citep{hastie2009elements}. 
Fourth, we develop a novel theory for DeFuSE, establishing its consistency for discovering the underlying DAG structure. 
DeFuSE requires an assumption for consistent causal discovery, called the \emph{strong causal minimality}, which is an analogy of the strong faithfulness \citep{uhler2013geometry} and the beta-min condition \citep{meinshausen2006high}. A central message of this paper is that nonlinearity plays an important role in causal discovery, permitting the separation of the nonlinear causal effects from linear confounding effects.

The rest of the article is structured as follows. 
Section \ref{sec:model} introduces the DAG model with hidden confounders and the proposed method DeFuSE. 
Section \ref{sec:fnn} implements DeFuSE based on feedforward neural networks for scalable computation.
Section \ref{sec:theory} provides a theoretical guarantee of DeFuSE for consistent discovery. 
Section \ref{sec:numerical-example} presents some numerical examples and compares DeFuSE with CAM, NOTEARS, {RFCI, and LRpS-GES}, followed by a discussion in Section \ref{sec:discussion}. 
The Appendix contains additional theoretical results and implementation details,
and the Supplementary Materials contain the technical proofs.

\section{Directed acyclic graph with confounders}\label{sec:model}

Consider a random vector $Y = (Y_1,\ldots,Y_p)$ {generated from a nonlinear 
structural equation model with additive confounders and noises},
\begin{equation}\label{model:confounding}
\begin{split}
Y_j = f_j\left(Y_{\pa(j)}\right)+ \eta_j + e_j,  \quad j\in V= \{1,\ldots,p\}, 
\end{split}
\end{equation}
where $f_j$ maps the subvector $Y_{\pa(j)} = (Y_k)_{k\in \pa(j)}$ to a real number, 
$\pa(j) \subseteq V\setminus \{ j\}$ 
is an index subset, $\eta = (\eta_1,\ldots,\eta_p) \sim N_p(0, \Sigma_\eta)$ 
is a vector of hidden confounders and is independent of random errors
$e=(e_1,\ldots,e_p)\sim N_p(0,\diag(\sigma_1^2,\ldots,\sigma_p^2))$, $\Sigma_\eta$ is an unknown covariance matrix, and $\diag(\sigma_1^2,\ldots,\sigma_p^2)$ is an unknown diagonal matrix. 
Then \eqref{model:confounding} is associated with a directed graph $G = (V, E)$ such that $E = \{ k\to j : k \in \pa(j), \ j \in V \}$. 
In this situation, $\pa(j)$ denotes the set of parents of $j$. 
Throughout this article, we assume that $G$ is a \emph{directed acyclic graph} (DAG) in that no directed path $j\to \cdots \to j$ exists in $G$. 
As a result, \eqref{model:confounding} generalizes the nonlinear DAG without unmeasured confounders \citep{hoyer2008nonlinear, peters2014causal} 
and the linear DAG \citep{peters2014identifiability}. 

In \eqref{model:confounding}, we assume the \emph{causal minimality} to ensure that the effect of each parent is non-vanishing. 
In other words, we require $\pa(j) = \ARG(f_j)$; $j=1,\ldots,p$, where $\ARG(f_j)$ denotes the minimal argument set $B\subseteq \pa(j)$ 
such that the value of $f_j$ only depends on $Y_B=(Y_k)_{k\in B}$. In particular, if $f_j$ is a constant function, we have $\pa(j) = \ARG(f_j) = \emptyset$. When $\eta\equiv 0$ (no confounder), this definition agrees with the usual {causal minimality} condition \citep{pearl2009causality}, 
requiring that the probability distribution of $Y$ is not Markov to any proper subgraph of $G$. The \emph{causal minimality}, as a form of causal 
faithfulness \citep{spirtes2000causation}, ensures that the problem of nonlinear causal discovery is well-defined. 

Equivalently, we rewrite \eqref{model:confounding} by letting $\varepsilon_j=\eta_j+ e_j$, 
\begin{equation}\label{model:dependence}
\begin{split}
Y_j=f_j\left(Y_{\pa(j)}\right) + \varepsilon_j, \quad j\in V = \{ 1,\ldots,p \},
\end{split}
\end{equation}
where $\varepsilon=(\varepsilon_1,\ldots,\varepsilon_p) \sim N(0,\Sigma)$ and $\Sigma=\Sigma_\eta + \diag(\sigma_1^2,\ldots,\sigma_p^2)$. 
Whereas \eqref{model:confounding} has a clear causal interpretation, 
\eqref{model:dependence} is simpler for the subsequent discussion. 
Our goal is to discover the causal relations between variables 
$Y_1,\ldots,Y_p$ by identifying $\{f_j\}_{1\leq j\leq p}$ and $\{\pa(j)\}_{1\leq j\leq p}$.
One major challenge is that the error $\varepsilon_j$ may be correlated with $Y_{\pa(j)}$ due to unmeasured confounders.

\subsection{Model identifiability}

\begin{figure}[ht]
\centering
\includegraphics[width=.25\textwidth]{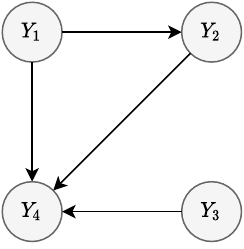}
\caption{Topological depth:
$\topdep_1 = \topdep_3 = 0$ (nodes $1$ and $3$ are root nodes),
$\topdep_2 = 1$, $\topdep_4=2$.
Here $\vx(1)=\{1,3\}$, $\vx(2)=\{1,2,3\}$, and $\vx(3) = V =\{1,2,3,4\}$.}
\label{fig:topological}
\end{figure}

This subsection establishes the identifiability conditions for \eqref{model:dependence}.
First, we introduce the concept of \emph{topological depth} for a DAG $G=(V,E)$ with nodes $V=\{1,\ldots,p\}$ and directed edges $E\subseteq V\times V$.
A node $j$ is a \emph{root} if it has no parent, i.e., $\pa(j)=\emptyset$. 
If there exists a directed path $k\to\cdots\to j$, then node $k$ is an ancestor of $j$ and $j$ is a descendant of $k$.
The \emph{topological depth} $\topdep_j$ of node $j\in V$ is the maximal length of 
a directed path from a root to $j$. Clearly, a root node has depth zero, and we have  $0\leq \topdep_j\leq d_{\max} \leq p-1$ for $j\in V$, where $d_{\max}$ is the length of the longest directed path in $G$. Let $\vx(d) = \{j : \topdep_j < d \}$ be the set of nodes with topological depth less than $d$, 
where $1\leq d\leq d_{\max} + 1$. 
Then $\emptyset\equiv \vx(0)\subseteq \vx(1)\subseteq\cdots\subseteq \vx({d_{\max}+1})=V$ and $\vx(\topdep_j)$ contains all the ancestors (and hence all the parents) of $Y_j$ but contains no descendant of $Y_j$. 
See Figure \ref{fig:topological} for an illustration. 

Next, we present a new condition for $\{f_j\}_{1\leq j\leq p}$ and $\{\pa(j)\}_{1\leq j\leq p}$ in \eqref{model:dependence} to be identifiable. 
For continuous function $f: \mathbb R^{m} \to \mathbb R$, $f$ is of \emph{sublinear growth} if $\lim_{\|x\|\to\infty} f(x)/\|x\| = 0$, where $\|\cdot\|$ is the Euclidean norm.

\begin{condition}\label{cond:sublinear}
Assume that $\{f_j\}_{1\leq j\leq p}$ are of sublinear growth. 
\end{condition}

For example, Condition \ref{cond:sublinear} is satisfied if $\{f_j\}_{1\leq j\leq p}$ are continuous and bounded. This sublinear growth assumption imposes restrictions on the nonlinearity of $\{f_j\}_{1\leq j\leq p}$, in contrast to the third-order differentiability condition for DAGs without confounders \citep{hoyer2008nonlinear,peters2014causal}. 

\begin{theorem}[Identifiability]\label{thm:identifiability}
Assume Condition \ref{cond:sublinear} is satisfied.
\begin{enumerate}[(A)]
  \item The sets $\vx(1)\subseteq\cdots\subseteq \vx({d_{\max}})$ are uniquely identifiable for almost every positive definite $\Sigma$ with respect to the Lebesgue measure, where the set of such $\Sigma$ is denoted as $\Psi$. Moreover, for $\Sigma\in\Psi$, if $\topdep_j=d$, then $Y_j-\E\left(Y_j\mid Y_{\vx(d)}\right)$ is normally distributed with mean zero and constant variance $\Var\left(Y_j\mid Y_{\vx(d)}\right)$; if $\topdep_j>d$, then $Y_j-\E\left(Y_j\mid Y_{\vx(d)}\right)$ is not normally distributed; $j=1,\ldots,p$.
  \item Given $\vx(1)\subseteq\cdots\subseteq \vx({d_{\max}})$, we have $\{f_j\}_{1\leq j\leq p}$ and $\{\pa(j)\}_{1\leq j\leq p}$ are well-defined and identifiable from the distribution of $Y$.
\end{enumerate}
\end{theorem}

By Theorem \ref{thm:identifiability}, model \eqref{model:dependence} is generically identifiable under Condition \ref{cond:sublinear}. 
Different from \citet{frot2019robust}, Theorem \ref{thm:identifiability} does not require pervasive confounding.
The sublinear growth assumption (Condition \ref{cond:sublinear}) allows us to separate the linear confounding effect from nonlinear causal relationships. 

\subsection{DeFuSE}

This subsection proposes the causal discovery method Deconfounded Functional Structure Estimation (DeFuSE). 
We commence with least squares regressions of $\{Y_j\}_{j\notin\vx(d)}$ on $Y_{\vx(d)}$,
\begin{equation*}
    Y_j = \underbrace{\E(Y_j \mid Y_{\vx(d)})}_{\text{(i)}} + \underbrace{Y_j - \E(Y_j \mid Y_{\vx(d)})}_{\text{(ii)}},
\end{equation*}
where (i) is the regression function and (ii) is the residual of the regression.
By Theorem \ref{thm:identifiability}, (ii) is normally distributed if and only if $d_j = d$, suggesting that normality tests (e.g. the Anderson-Darling test \citep{anderson1952asymptotic}) for $\{Y_j - \E(Y_j \mid Y_{\vx(d)})\}_{j\notin \vx(d)}$ can be utilized to identify $V(d+1)$. 
Further, if $d_j = d$, then (i) becomes
\begin{equation*} 
\begin{split}
\E(Y_j \mid Y_{\vx(d)})=f_j(Y_{\pa(j)})+\E(\varepsilon_j \mid Y_{\vx(d)}),
\end{split}
\end{equation*}
where $\E(\varepsilon_j \mid Y_{\vx(d)})$ is the bias arising from hidden confounding.
Theorem \ref{deconfounding} allows us to estimate $\{f_j\}_{j\in\vx(d+1)}$ and $\{\pa(j)\}_{j\in\vx(d+1)}$ by regressions with deconfounding adjustment. 

\begin{theorem}\label{deconfounding} 
In \eqref{model:dependence}, if $\topdep_j = d$, then
\begin{equation}\label{conditional-mean}
\begin{split}
\E(Y_j \mid Y_{\vx(d)} )
= f_j(Y_{\pa(j)}) + \left\langle \xi_{\vx(d)}, \beta_j\right\rangle,
\end{split}
\end{equation}
where $\xi_{\vx(d)} \equiv ( Y_k-\E(Y_k \mid Y_{\vx(\topdep_k)}))_{k\in\vx(d)}$, 
$\beta_j$ is a parameter vector, 
$\langle\cdot,\cdot\rangle$ is the Euclidean inner product, and we define $\left\langle \xi_{\vx(d)}, \beta_j\right\rangle\equiv 0$ whenever $V(d) = \emptyset$.
\end{theorem}

Now, we develop an algorithm that iteratively estimates $\vx({d+1})$, $\xi_{\vx({d+1})}$, $\{f_j\}_{j\in \vx(d+1)}$, and $\{\pa(j)\}_{j\in \vx(d+1)}$, given $\vx(d)$ and $\xi_{\vx(d)}$ as input. 
To proceed, suppose an independent sample $\{(Y_{1}^{(i)},\ldots, Y_{p}^{(i)})\}_{1\leq i\leq n}$ from model \eqref{model:dependence} is given.
Let $\widehat{\xi}_{\vx(d)}^{(i)}
= (Y_{k}^{(i)} - \widehat{Y}_{k}^{(i)})_{k\in \vx(d)}$ 
be the estimated residual vector for the $i$-th observation,
where $\widehat{Y}_{k}^{(i)}=\widehat{f}_k\big(Y_{\vx(\topdep_k)}^{(i)}\big) + \big\langle\widehat{\xi}_{\vx(\topdep_k)}^{(i)},\widehat{\beta}_j \big\rangle$ for $k \in \vx(d)$.
Based on \eqref{conditional-mean}, we regress each variable in $\{Y_j\}_{j\notin \vx(d)}$ on $\left(Y_{\vx(d)},\xi_{\vx(d)}\right)$, 
\begin{equation}
\label{regression}
\begin{split}
(\widehat{f}_j,\widehat{\beta}_j) = 
\argmin_{\{ (f_j,\beta_j) : f_j \in\mathcal F_j\}} \
\sum_{i=1}^n
\Big(Y_{j}^{(i)}- f_j\big( Y_{\vx(d)}^{(i)}\big) - \big\langle \widehat{\xi}_{\vx(d)}^{(i)},\beta_j \big\rangle \Big)^2 
 \quad \text{s.t.} \quad |\ARG(f_j)| \leq \kappa_j,
\end{split}
\end{equation}
where $|\ARG(f_j)|$ is the effective input dimension of $f_j$, $\kappa_j \geq 0$ is an integer-valued hyperparameter and is estimated via a standalone validation set (see Section \ref{sec:implementation}), and $\mathcal{F}_j$ is a function space consisting of sublinear growth continuous functions. 
Then we perform normality tests for $\{(\widehat{\xi}_{j}^{(1)}, \ldots, \widehat{\xi}_{j}^{(n)})\}_{j\notin\vx(d)}$, and estimate $\vx(d+1)$ by including $\vx(d)$ and all the indices failing to reject the tests.
Finally, we estimate $\{\widehat{\pa}(j)\}_{j\in\vx(d+1)}$ by $\widehat{\pa}(j) = \ARG(\widehat{f}_j)$. 

We summarize the procedure in Algorithm \ref{algorithm:defuse},
where a bold-face letter denotes a data vector/matrix of sample size $n$.

\begin{algorithm}
  \caption{DeFuSE} \label{algorithm:defuse}
  \KwIn{An $n\times p$ data matrix $\bm Y = (\bm Y_1,\ldots, \bm Y_p)$; }
  \Parameter{significance level $\alpha$ for normality test; hyperparameters $\{\kappa_j\}_{1\leq j\leq p}$;}

  Let $V(0)\leftarrow \emptyset$ and $d\leftarrow 0$\;
  \While{$\vx(d)\neq V$}{
        Regress $\{\bm Y_j\}_{j\notin \vx(d)}$ on $(\bm Y_{\vx(d)},\widehat{\bm\xi}_{\vx(d)})$ based on \eqref{regression}\;

        Update $\{\widehat{\bm \xi}_{j} \leftarrow \bm Y_j - \widehat{\bm Y}_j \}_{j \notin \vx(d)}$\;

        Let $\vx({d+1})\leftarrow \vx(d)\cup\{ j \notin \vx(d) : \widehat{\bm \xi}_j \text{ fails to reject the normality test} \}$\;

        Let $\{\widehat{\pa}(j) \leftarrow \ARG(\widehat{f}_j)\}_{j\in\vx(d+1)}$ and $d\leftarrow d+1$\;
  }
  \KwOut{$\{\widehat{f}_j\}_{1\leq j\leq p}$ and $\{\widehat{\pa}(j)\}_{1\leq j\leq p}$;}
\end{algorithm}

\begin{remark}[Normality test and the choice of $\alpha$]\label{remark:normal-test}
For implementation, we use the Anderson-Darling test \citep{anderson1952asymptotic} to examine the null hypotheses 
\begin{equation*}
    \mathcal{H}_0^{(j,d)} : Y_j - \E(Y_j\mid Y_{V(d)}) \text{ is normal}; \quad j\notin V(d), \quad 0\leq d\leq d_{\max}.
\end{equation*}
Other tests or metrics, such as the Wasserstein distance, can also be used. Moreover, the normality test can be combined with a goodness-of-fit measure to further improve performance. The significance level $0<\alpha<1$ is a hyperparameter similar to that in the PC algorithm \citep{kalisch2007estimating}.
To choose $\alpha$, denoting by $\mathcal T$ the set of true null hypotheses, then  
$P\Big(\text{some }\mathcal H_0^{(j,d)}\in\mathcal T \text{ is rejected}\Big)
    \leq \sum_{\mathcal H_0^{(j,d)}\in\mathcal T} P\Big(\mathcal H_0^{(j,d)} \text{ is rejected}\Big) \approx |\mathcal T|\alpha$.
For $1\leq d\leq d_{\max}+1$, identifying $\vx(d)$ requires $p-|\vx(d-1)|$ tests, among which $|\vx(d)|-|\vx(d-1)|$ null hypotheses are true and $p-|\vx(d)|$ are not. Thus, $|\mathcal T| = \sum_{d=1}^{d_{\max}+1}(|\vx(d)|-|\vx(d-1)|) = p$, suggesting an empirical rule $\alpha = o(1/p)$ so that $|\mathcal T|\alpha \to 0$.
\end{remark}

Finally, Example \ref{example:identifiability} illustrates
the importance of deconfounding for causal discovery.

\begin{figure}[ht]
  \centering
  \includegraphics[width=.75\textwidth]{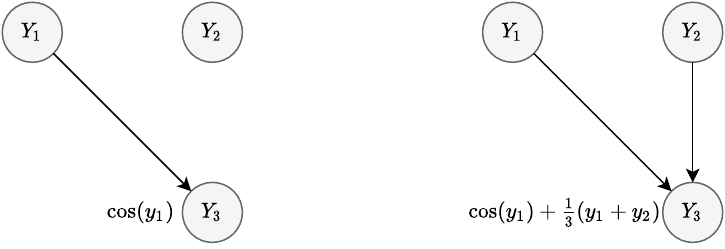}
  \caption{Display of the directed acyclic graph in Example \ref{example:identifiability}.}
  \label{fig:identifiability}
\end{figure}

\begin{example} \label{example:identifiability}
Consider a special case of \eqref{model:confounding} with three variables,
\begin{equation}\label{ex1}
\begin{split}
Y_1 = e_1 + \eta, \quad
Y_2 = e_2 + \eta, \quad
Y_3 = \cos(Y_1) + e_3 + \eta,
\end{split}
\end{equation}
where $e_1,e_2,e_3,\eta\sim N(0,1)$ independently; see Figure \ref{fig:identifiability}. 
As a special case of \eqref{conditional-mean}, we have $\E(Y_3 \mid Y_1,Y_2) = \cos(Y_1)+ \E (\eta\mid Y_1,Y_2)=\cos(Y_1) + Y_1/3 + Y_2/3$, where $\topdep_3 = 1$, $\vx(1) = \{1,2\}$, $\xi_{\vx(1)} = (\xi_1,\xi_2) = (e_1+\eta,e_2+\eta)$, and $\xi_{\vx(2)} = \xi_3 = e_3 + (\eta - e_1 - e_2)/3$.
The presence of $Y_2/3$ is due to the confounder $\eta$.
If we have regressed $Y_3$ on $Y_1$ and $Y_2$ to identify the parent variables of $Y_3$, 
then the regression would yield a true discovery $Y_1\to Y_3$ and a false discovery $Y_2\to Y_3$. Consequently, direct regression of $Y_j$ on $Y_{\vx(\topdep_j)}$ without any adjustment renders false discovery of functional causal relations.
\end{example}

\section{DeFuSE via neural networks}\label{sec:fnn}

Solving \eqref{regression} is challenging for a large-scale problem due to fitting nonparametric functions. Existing nonparametric methods such as tensor-product splines and kernels are not scalable in a growing sample size and dimension. 
For example, tensor-product B-splines least squares regression suffers from exponential growth of time and space complexity with increasing dimensions. To overcome this difficulty, we solve \eqref{regression} via a feedforward neural network (FNN) together with stochastic gradient descent for scalable computation.

Specifically, for $\topdep_j \geq d$, we approximate 
$f_j\left(Y_{\vx(d)}\right) + \left\langle \xi_{\vx(d)},\beta_j\right\rangle$ by an FNN,
\begin{equation}
\label{deep0}
g_j\left(Y_{\vx(d)}, \xi_{\vx(d)}\right) 
= f_j^L\circ \cdots \circ f_j^1\left(Y_{\vx(d)}\right) + \left\langle \xi_{\vx(d)},\beta_j\right\rangle, 
\ f^l_j(\cdot) = \sigma^l\left(W^l (\cdot) + b^l\right); \ l = 1,\ldots,L,
\end{equation} 
where $W^l\in\mathbb{R}^{h_{l}\times h_{l-1}}$ is the weight matrix of links from the $(l-1)$-th to the $l$-th layer, $b^l\in\mathbb{R}^{h_l}$ is the bias vector in the $l$-th layer, $h_l$ is the number of neurons in the $l$-th layer with $h_l=h$; $l=1,\ldots,L-1$, and $h_L=1$, $L$ is the number of layers, and $\sigma^l(\cdot)$ is an activation function. For $l=1,\ldots,L-1$, we use the Rectifier Linear Unit (ReLU) activation $\sigma^l(z)=\max(0,z)$.

To solve \eqref{regression}, consider a FNN parameter vector $\theta_j=((W^l_j, b^l_j)_{1\leq l \leq L}, \beta_j)$ which belongs to a parameter space $\Theta_d$.   
We impose constraints $\sum_{k\in \vx(d)}
\min(\|W^{1}_k \|/\tau,1) \leq \kappa_j$
on the $k$-th column $W^{1}_k$ of the weight matrix $W^1$ at the first layer to enforce the constraint $|\ARG(f_j)|\leq \kappa_j$ in \eqref{regression},
where $\min(|\cdot|/\tau,1)$ is to approximate $\I(\cdot \neq 0)$ as $\tau \rightarrow 0^+$ \citep{shen2012likelihood}. As such, if $W^{1}_{k}=0$ then $g_j\left(Y_{\vx(d)},\xi_{\vx(d)}\right)$ does not depend on $Y_k$. 
Finally, we regularize the FNN by an $L_2$-norm constraint $\|\theta_j\|\leq s$ on the model parameters $\theta_j$ for numerical consideration. 
This leads to the following regression for estimating $(f_j, \beta_j)$,
\begin{equation}
\label{deep}
\begin{split}
\min_{\{\theta_j:\|\theta_j\|\leq s\}} \quad& \
\sum_{i=1}^n \Big(Y_{j}^{(i)} - f_j\big(Y_{\vx(d)}^{(i)}\big) - \big\langle \widehat{\xi}_{\vx(d)}^{(i)}, \beta_j\big\rangle \Big)^2, \\
\quad\text{s.t.} \qquad&
\sum_{k\in \vx(d)}
\min(\|W^{1}_k \|/\tau,1)\leq \kappa_j, \quad 
\sum_{k\in\vx(d)} \min(|\beta_{j,k}|/\tau,1)\leq \varsigma_j,
\end{split}
\end{equation}
where $\tau>0$, $0\leq \kappa_j\leq |\vx(d)|$, $0\leq \varsigma_j\leq |\vx(d)|$, and $s \geq 0$ are hyperparameters. See Section \ref{sec:implementation} for more details on network training and hyperparameter tuning.

\begin{remark}
  Algorithm \ref{algorithm:defuse} requires $O(d_{\max} (p-1))$ normality tests and regressions \eqref{regression}. Each regression \eqref{regression}, solved by \eqref{deep} with stochastic gradient descent, requires $O(N_\textnormal{epoch} n \dim(\theta) )$ operations, where $N_{\textnormal{epoch}}$ is the number of epochs in training and one epoch means that each sample in training has an opportunity to update model parameters.  
\end{remark}

\section{Learning theory}\label{sec:theory}

This section develops a novel theory to quantify the finite-sample error of DeFuSE.
In what follows, $c_1$-$c_6$ are positive constants and $^\circ$ decorates the truth. Let $\mathcal G_j$ be the function space of regression functions $g_j(\cdot, \star) = f_j(\cdot) +  \left\langle \star, \beta^\circ_j\right\rangle$, and denote the true regression function by $g^\circ_j(\cdot,\star)= f_j^\circ(\cdot) + \left\langle \star, \beta^\circ_j \right\rangle$. 
By definition, ${\pa}^\circ(j) = \ARG(f^\circ_j)$.

\begin{condition}\label{cond:approximation} 
  There exists an approximating function $g_j^*(\cdot, \star) = f_j^*(\cdot) +  \left\langle \star, \beta^\circ_j\right\rangle\in\mathcal G_j$
  such that $\|g^*_j- g^\circ_j\|_{L_2} = \| f^*_j - f^\circ_j\|_{L_2} \leq c_3\epsilon_n$;
  $j=1,\ldots,p$, where $\|\cdot\|_{L_2}$ is the $L_2$-norm with respect to measure $P$. Moreover, assume $\{f_j^\circ\}_{1\leq j\leq p}$ are continuous and $\|f^\circ_j\|_{\infty}\leq c_1$, where $\|\cdot\|_\infty$ is the sup-norm. 
\end{condition}

To measure the signal strength, we define the degree of nonlinear separation as 
\begin{equation*}
  D_{\min}=\min_{1\leq j\leq p}\inf\left\{ \frac{\|g_j - g^\circ_j\|_{L_2}^2} {|\pa^\circ(j)\setminus \ARG(f_j)|} : 
\begin{aligned}
  g_j\in\mathcal{G}_j, \qquad \ARG(f_j)\neq \pa^\circ(j),\\
  \|\beta_{j}\|_0 \leq \varsigma^\circ, \ |\ARG(f_j)|\leq |\pa^\circ(j)| 
\end{aligned}\right\}.
\end{equation*}
\begin{condition}[Strong causal minimality]\label{cond:separation}
Assume $D_{\min} \geq c_4 \max\big(4 \epsilon^2_n, n^{-1} \log n, n^{-1} \log p\big)$,
where $c_4 \geq 1$.
\end{condition}

The strong causal minimality (Condition \ref{cond:separation}) requires that the signal strengths of parent 
variables are sufficiently strong so that the corresponding causal function is distinguishable from those 
supported on non-parent variables. It is a strong version of the causal minimality for nonlinear causal discovery from a finite sample, similar to the strong faithfulness \citep{uhler2013geometry} for linear causal discovery and the beta-min condition \citep{meinshausen2006high} for high-dimensional variable selection.

\begin{theorem}[Error bounds for DeFuSE]\label{thm:variable-selction} 
Assume Conditions \ref{cond:sublinear}-\ref{cond:separation}, Conditions \ref{cond:entropy}-\ref{cond:confounding} in Section \ref{sec:regularity-conditions} are met and $\Sigma\in\Psi$. 
\begin{enumerate}[(A)]
  \item The DAG recovery error is 
  $P(\widehat{G} \neq G^\circ) \leq c_6\exp(-c_5n\epsilon_n^2 - \log n) + \pi_{\alpha}(G^\circ)$,
  when the hyperparameters $\kappa_j=|{\pa}^\circ(j)|$ and $\|\beta_j^\circ\|_0\leq \varsigma_j \leq \varsigma^\circ$; $1\leq j\leq p$, where $\pi_{\alpha}(G^\circ)$ is the normality test error given the true model.
  Consequently, $P(\widehat{G} \neq G^\circ) \rightarrow 0$ 
  provided that $\pi_{\alpha}(G^\circ)\to 0$, as $n \rightarrow \infty$.

  \item The regression estimation error is 
  $\max_{1\leq j\leq p} \|\widehat{g}_j - g^\circ_j\|_{L_2}=O_p(\epsilon_n)$. 
  Suppose $f^\circ_j$ satisfies $\| f^\circ_j \|_\infty\leq C$ and has bounded support; $1\leq j\leq p$. Then the causal function estimation error is 
  $\max_{1\leq j\leq p} \|\widehat{f}_j - f^\circ_j\|_{L_2}= O_p(\epsilon_n)$
  provided that $\|\widehat{f}_j\|_\infty \leq C'$ for $C'\geq C$.
\end{enumerate}
\end{theorem}

Typically, we have $\pi_{\alpha}(G^\circ)\to 0$ when $\alpha = o(1/p)$ and the dimension $p$ does not grow too fast. Moreover, Theorem \ref{thm:variable-selction} indicates that hyperparameter $\kappa_j$ is critical to consistent discovery, while $\varsigma_j$ is less important provided that $\varsigma_j\geq\|\beta_j^\circ\|_0$ and is not too large; see also Section \ref{sec:implementation}.

Next, we apply Theorem \ref{thm:variable-selction} to the implementation via FNNs in \eqref{deep}. Before proceeding, we define $\mathcal C^{r}_j$, the space of functions with $r$-continuous derivatives over the domain $\mathbb{R}^{|\pa^\circ(j)|}$. For any function $f_j \in \mathcal C^{r}_j$, the $\mathcal C^{r}_j$-norm of $f_j$ is defined as 
\begin{equation*}
    \|f_j\|_{\mathcal C^r_j}=\sum_{\bm\alpha:|\bm\alpha|<r} \|\partial^{\bm\alpha}f_j\|_{\infty}
  +\sum_{\bm\alpha:|\bm\alpha|=\lfloor r\rfloor} \sup_{x_1\neq x_2}
  \frac{|\partial^{\bm\alpha} f(x_1)-\partial^{\bm\alpha} f(x_2)|}{\|x_1-x_2\|^{r-\lfloor r\rfloor}_{\infty}},
\end{equation*}
where $\partial^{\bm\alpha}=\partial^{\alpha_1}\cdots\partial^{\alpha_{|\pa(j)|}}$ with $\bm\alpha\in\mathbb{N}^{|\pa(j)|}$ and $|\bm\alpha|=\sum_{k=1}^{|\pa(j)|}\alpha_k$; $j=1,\ldots,p$.
In what follows, $C_1$-$C_3$ are positive constants that may depend on $(\kappa^\circ,r)$. 

\begin{condition}
\label{cond:fnn}
  Assume $f^\circ_j\in \left\{ f_j\in \mathcal C^r_j : \|f_j\|_{\mathcal C^r_j} \leq C_1\right\}$, 
  where $r$ does not depend on $(p,n)$.
\end{condition}

\begin{theorem}[Consistency of FNN-DeFuSE] \label{thm:fnn} 
Under Conditions \ref{cond:separation}-\ref{cond:fnn}, and \ref{cond:confounding} in Section \ref{sec:regularity-conditions}, DeFuSE implemented by FNNs in \eqref{deep} consistently recovers all causal relations defined in \eqref{model:dependence} with $\epsilon_n^2 =C_3 (n^{-r/(r + \kappa^\circ + \varsigma^\circ)}(\log n)^3 +n^{-1}(\kappa^\circ+\varsigma^\circ)\log p)$ in Theorem \ref{thm:variable-selction}, provided that the width of the FNN $h=C_2\epsilon_n^{-\kappa^\circ/r}$ and its depth $L=C_2 \log(1/\epsilon_n)$, the hyperparameters $s= C_2\epsilon_n^{-(\kappa^\circ+\varsigma^\circ)/r}\log(1/\epsilon_n)$, $\kappa_j=|\pa^\circ(j)|$, $\|\beta_j^\circ\|_0\leq \varsigma_j \leq \varsigma^\circ$; $j=1,\ldots,p$. 
Here, the FNN function space $\mathcal{G}_j = \{g_j = g_j(\cdot;\theta) : \theta\in \Theta_j\}$ is associated with the FNN parameter space
\begin{equation*}
      \Theta_j=
      \left\{ \theta = ((W^l,b^l)_{1\leq l \leq L},\beta_j): 
    \ \max_{1\leq l\leq L}h_l\leq h, \ \|\theta\| \leq s \right\}; \quad j=1,\ldots,p.
\end{equation*}
\end{theorem}

It is worth noting that the rate $\epsilon_n^2\asymp n^{-r/(r + \kappa^\circ + \varsigma^\circ)} (\log n)^3 + n^{-1}(\kappa^\circ+\varsigma^\circ) \log p$ for FNN relies on the approximation result of \citet{schmidt2019deep} as well as the choice of $L$, $h$, and $s$. This rate agrees with \citet{farrell2021deep} up to logarithm terms; however, it is slower than $n^{-r/(r + (\kappa^\circ+\varsigma^\circ)/2)}$ in view of \citet{stone1982optimal} for nonparametric regression over $[0,1]^{\kappa^\circ+\varsigma^\circ}$, suggesting that it may be suboptimal. This may be due to the approximation, namely the use of non-differentiable ReLU FNNs to approximate smooth functions.

\section{Numerical examples}\label{sec:numerical-example}

\subsection{Simulations}

This subsection examines the operating characteristics of DeFuSE and compares DeFuSE with CAM \citep{buhlmann2014cam}, NOTEARS (FNN version) \citep{zheng2020learning}, LRpS-GES \citep{frot2019robust}, and RFCI \citep{colombo2012learning}. We implement DeFuSE in Python. For competitors, we use R packages for CAM (\texttt{CAM}), RFCI (\texttt{pcalg}), and LRpS-GES (\texttt{lrpsadmm} and \texttt{pcalg}), and use a Python program for NOTEARS (\texttt{notears}).

In simulations, we consider two types of DAGs with hidden confounders. Define an adjacency matrix $\bm U=(U_{jk})_{p \times p}$ of a DAG as $U_{jk}=1$ if $j \in \pa(k)$ and $0$ otherwise.

\paragraph{Random DAG.}
Consider a sparse graph where the edges are added independently with equal probability. 
In particular, an adjacency matrix $\bm U\in \{0,1\}^{p\times p}$ is randomly generated: $P(U_{jk} = 1) = s$ if $j < k$ and $P(U_{jk} = 1) =0$ otherwise, where $s$ controls the degree of sparseness of the DAG. In our simulation, we choose $s = 1/p$. 

\paragraph{Hub DAG.}
Consider a sparse graph with a hub node. Let $\bm U\in\{0,1\}^{p\times p}$, 
where $U_{1k} =1$ and $U_{jk}=0$ otherwise. 
In this case, node 1 has a dense neighborhood, but the whole DAG remains sparse.

\paragraph{Simulated data.}
Given $\bm U$, we generate a random sample of size $n$ from 
\begin{equation}\label{example:structural-equation}
Y_j = \alpha_{0} Y_{k_1}Y_{k_2} + \sum_{k\in\pa(j)}\alpha_{j,k} f_{j,k}(Y_{k} + \omega_{j,k}) + \varepsilon_j; 
\quad j = 1,\ldots,p,
\end{equation}
where the function $f_{j,k}$ is randomly sampled from $\{ x\mapsto x^2, x\mapsto\cos(x) \}$, the coefficients $\alpha_{j,k} \sim \text{Uniform}([-3,-2]\cup[2,3])$, $\omega_{j,k} \sim \text{Uniform}([-1,1])$, and 
\begin{equation*}
  \begin{cases}
    \alpha_0 = 0, & |\pa(j)| = 1,\\
    \alpha_0 = 1, \ k_1,k_2 \text{ are randomly sampled from } \pa(j), & |\pa(j)|>1.
  \end{cases}
\end{equation*}
For error terms, let $\varepsilon\sim N(0,\Sigma)$ with 
$\Sigma_{jj} = 2$ for $1\leq j\leq p$, $\Sigma_{2k-1,2k} = \Sigma_{2k,2k-1} = 1$ for $1 \leq k\leq \lfloor p/2\rfloor$, and $\Sigma_{jj'}=0$ otherwise.
Of note, \eqref{example:structural-equation} violates Condition \ref{cond:sublinear} as the functions $(y_1,y_2)\mapsto \alpha_0 y_1y_2$ and $f_{j,k}$ may not be of sublinear growth. 

\paragraph*{Metrics.}
For evaluation, we consider four graph metrics: 
the false discovery rate (FDR), 
the false positive rate (FPR), 
the true positive rate (TPR),
and the structural Hamming distance (SHD). 
To compute the metrics, let TP, RE, and FP be the numbers of identified 
edges with correct directions, those with wrong directions, and estimated edges not in the skeleton of the true graph. 
Moreover, denote by PE the total number of estimated edges, 
TN the number of correctly identified non-edges,  
and FN the number of missing edges compared to the true skeleton. 
Then
\begin{align*}
\textrm{FDR} &= (\textrm{RE} + \textrm{FP})/\textrm{PE}, & 
\textrm{FPR} &= (\textrm{RE} + \textrm{FP})/(\textrm{FP} + \textrm{TN}), \\
\textrm{TPR} &= {\textrm{TP}}/ 
              {(\textrm{TP} + \textrm{FN})}, &
\textrm{SHD} &= \textrm{FP} + \textrm{FN} + \textrm{RE}.
\end{align*}
Note that LRpS-GES outputs a completed partially DAG (CPDAG) 
and RFCI outputs a partial ancestral graph (PAG). 
Both PAG and CPDAG may contain undirected edges, in which case they are evaluated favorably by assuming the correct directions for undirected edges whenever possible, similar to \citet{zheng2020learning}.

\begin{sidewaystable}[ph!] 
\footnotesize
\centering
\caption{Averaged false positive rate (FPR), false discovery rate (FDR), true positive rate (TPR), structural Hamming distance (SHD), and their standard deviations in parenthesis, for five methods based on 50 replications. A smaller value of FPR, FDR, and SHD indicates higher accuracy, whereas a larger value of TPR means higher accuracy. For DeFuSE*, the data are standardized. For hub DAG, when $p=100$ and $n=500$, LRpS-GES fails to deliver the computational results after 96 hours.}
\label{table:main}
\begin{tabular}{l l c c c c c c c c c c}
\hline
Graph      &            & \multicolumn{4}{c}{Random}                        & & \multicolumn{4}{c}{Hub}           \\       
\cline{3-6} \cline{8-11}
$(p,n)$    & Method     & FPR       & FDR        & TPR       & SHD          & & FPR        & FDR        & TPR       & SHD          \\ \hline
(30,500)   & DeFuSE     & .00 (.00) & .12  (.06) & .93 (.04) & ~~2.6 (~1.2) & & .00 (.00)  & .06 (.06)  & .87 (.10) & ~~5.3 (~4.6) \\ 
           & DeFuSE*    & .00 (.00) & .13  (.11) & .93 (.07) & ~~1.7 (~1.4) & & .00 (.00)  & .07 (.10)  & .91 (.16) & ~~4.2 (~5.6) \\
           & CAM        & .03 (.00) & .52  (.02) & 1.0 (.02) & ~14.2 (~1.0) & & .09 (1.0)  & .69 (.05)  & .53 (.07) & ~48.2 (~6.9) \\
           & NOTEARS    & .28 (.07) & .91  (.02) & .80 (.13) & 120.2 (31.6) & & .19 (.02)  & .84 (.05)  & .52 (.17) & ~94.3 (12.8) \\ 
           & RFCI       & .07 (.01) & .89  (.03) & .29 (.11) & ~26.8 (~1.2) & & .22 (.02)  & .95 (.01)  & .04 (.01) & ~74.4 (~3.7) \\ 
           & LRpS-GES   & .07 (.01) & .91  (.03) & .21 (.07) & ~31.9 (~1.7) & & .08 (.01)  & .92 (.01)  & .06 (.01) & ~44.5 (~1.4) \\ \hline
(100,500)  & DeFuSE     & .00 (.00) & .03  (.03) & .92 (.03) & ~~4.0 (~1.7) & & .00 (.00)  & .05 (.03)  & .72 (.24) & ~31.4 (23.7) \\
           & DeFuSE*    & .00 (.00) & .16  (.06) & .85 (.06) & ~10.6 (~3.0) & & .00 (.00)  & .10 (.18)  & .71 (.27) & ~32.9 (26.2) \\
           & CAM        & .01 (.00) & .61  (.01) & 1.0 (.01) & ~57.4 (~2.5) & & .05 (.01)  & .94 (.01)  & .16 (.03) & 306.3 (13.0) \\ 
           & NOTEARS    & .04 (.02) & .93  (.04) & .18 (.15) & 130.6 (24.8) & & .18 (.02)  & .96 (.01)  & .03 (.05) & 992.6 (65.4) \\ 
           & RFCI       & .02 (.00) & .95  (.02) & .15 (.06) & ~83.5 (~1.1) & & .07 (.01)  & .99 (.01)  & .01 (.00) & 268.6 (~6.7) \\ 
           & LRpS-GES   & .02 (.00) & .96  (.01) & .10 (.04) & ~83.3 (~2.0) & & -          &  -         &  -        & -            \\  
\hline
\end{tabular}
\end{sidewaystable}

As suggested in Table \ref{table:main}, DeFuSE performs the best across all the situations in terms of FPR, FDR, TPR, and SHD. As expected, CAM and NOTEARS cannot treat unobserved confounders, whereas RFCI and LRpS-GES cannot deal with nonlinear causal relationships.
It is worth noting that DeFuSE* takes standardized data as input and achieves comparable performance to DeFuSE, indicating that DeFuSE is insensitive to the degree of varsortability \citep{reisach2021beware}.
Moreover, DeFuSE seems robust in the absence of Condition \ref{cond:sublinear}; see also Theorem \ref{thm:alternative-identifiability} in Appendix and discussions there. Overall, nonlinearity helps identify causal relations, allowing for a separation of nonlinear causal effects from linear confounding effects.

\paragraph*{Sensitivity to normality test significance level $\alpha$.}
In the above experiments, we use the Anderson-Darling test \citep{anderson1952asymptotic} with $\alpha = 0.025$ as the default choice. Now, we assess the algorithmic sensitivity to different choices of $\alpha\in\{ 0.1, 0.05, 0.025, 0.01 \}$.   

As suggested in Table \ref{table:sensitivity}, the overall performance of DeFuSE seems insensitive to the choice of $\alpha$, although the default choice $\alpha=0.025$ may be sub-optimal. Based on our limited numerical experience, we suggest $\alpha = o(1/p)$ as an empirical rule to reduce the tuning cost of $\alpha$; see also Remark \ref{remark:normal-test}.

\begin{table}[ht] 
  \footnotesize
  \centering
  \caption{Sensitivity analysis: Averaged false positive rate (FPR), false discovery rate (FDR), true positive rate (TPR), structural Hamming distance (SHD), and their standard deviations in parenthesis, for different choices of $\alpha$ based on 50 replications. A smaller value of FPR, FDR, and SHD indicates higher accuracy, whereas a larger value of TPR means higher accuracy. Here, $p=30$ and $n=500$.}
  \label{table:sensitivity}
  \begin{tabular}{l l c c c c c}
  \hline
Graph       & $\alpha$   & FPR        & FDR        & TPR        & SHD       \\ \hline
Random      & .100       & .00 (.00)  & .12 (.08)  & .95 (.05)  & 2.4 (1.7) \\ 
            & .050       & .00 (.00)  & .13 (.07)  & .96 (.04)  & 2.4 (1.5) \\
            & .025       & .00 (.00)  & .12 (.06)  & .93 (.04)  & 2.6 (1.2) \\ 
            & .010       & .00 (.00)  & .13 (.07)  & .92 (.07)  & 3.0 (1.6) \\ \hline
Hub         & .100       & .00 (.00)  & .08 (.04)  & .91 (.04)  & 5.0 (2.5) \\ 
            & .050       & .00 (.00)  & .05 (.04)  & .95 (.03)  & 3.0 (2.0) \\
            & .025       & .00 (.00)  & .06 (.06)  & .87 (.10)  & 5.3 (4.6) \\ 
            & .010       & .00 (.00)  & .03 (.02)  & .97 (.02)  & 1.8 (1.5) \\ 
  \hline
  \end{tabular}
\end{table}

\subsection{Real data analysis}

This subsection applies DeFuSE to reconstruct gene regulatory networks for the Alzheimer's Disease Neuroimaging Initiative (ADNI) data. 
In particular, we construct two gene networks respectively for Alzheimer's Disease (AD) and healthy subjects to highlight some gene-gene interactions differentiating 
patients with AD/cognitive impairments and healthy individuals.

The ADNI dataset (\url{http://adni.loni.usc.edu/}) includes gene expressions, whole-genome sequencing, and phenotypic data. 
After cleaning and merging, we obtain a sample of 712 subjects 
in four groups, Alzheimer's Disease (AD), Early Mild Cognitive Impairment (EMCI), Late Mild Cognitive Impairment (LMCI), and Cognitive Normal (CN). 
For our purpose, we treat 247 CN individuals as controls while the remaining 465 individuals 
as cases (AD-MCI). Previous studies suggest that the amyloid precursor protein, 
the presenilin proteins, and the tau protein may involve in AD \citep{o2011amyloid,kelleher2017presenilin,palmqvist2020discriminative}, so we focus on the metabolic pathways of these proteins. 
Specifically, we extract the reference pathways in \url{https://genome.jp/pathway/map05010} from the KEGG database \citep{kanehisa2000kegg}, including $p = 20$ genes in the data. 

For data analysis, we first regress the gene expression levels on five covariates, Gender, Handedness, Education level, Age, and Intracranial volume, then use the residuals as gene expressions in the subsequent analysis. We normalize all gene expression levels and use the same FNN structure for fitting as in the simulation study. The normality test is conducted at a significance level $\alpha = 0.05$. 

\begin{figure}
  \centering
  \subfloat[AD-MCI]{\includegraphics[width=0.4\textwidth]{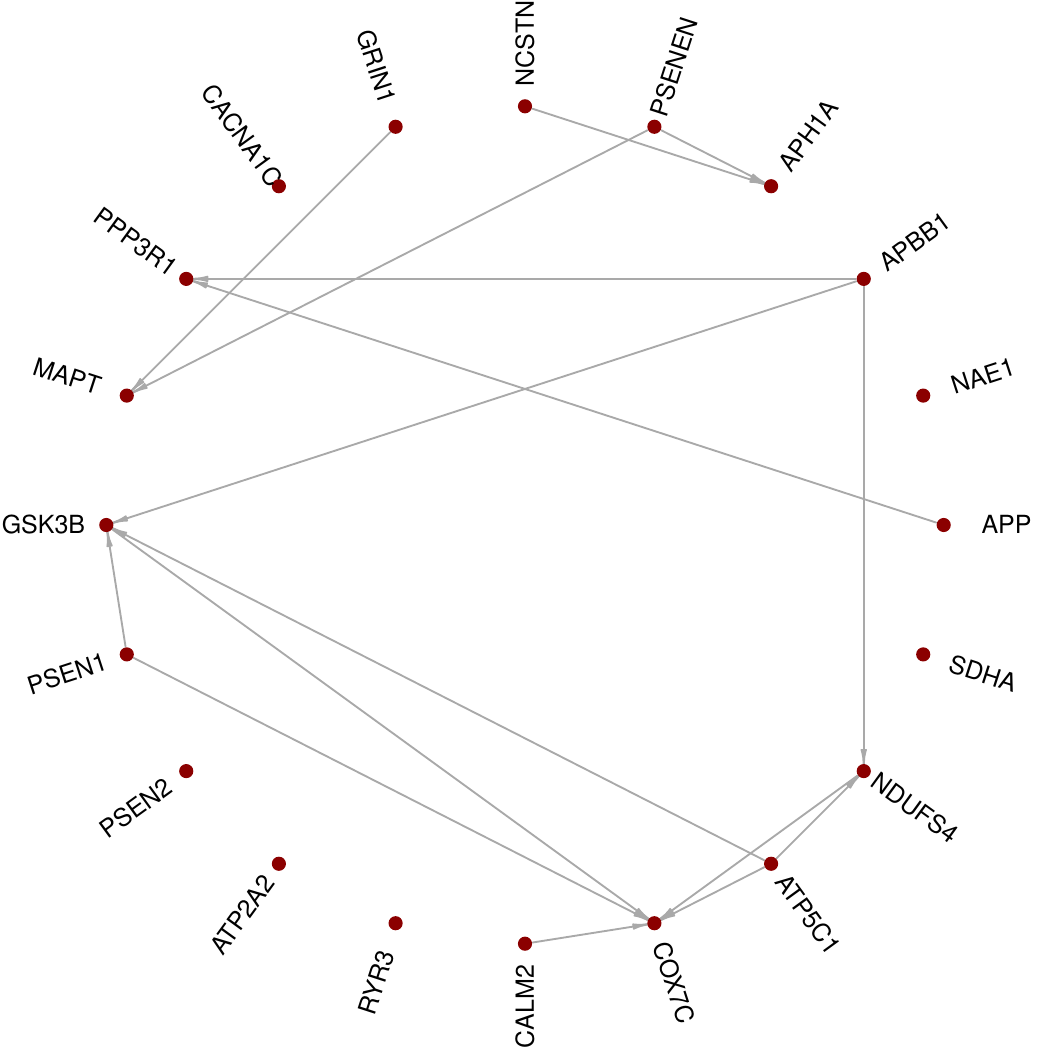}\label{fig:real-data-ad}}
  \qquad
  \subfloat[CN]{\includegraphics[width=0.4\textwidth]{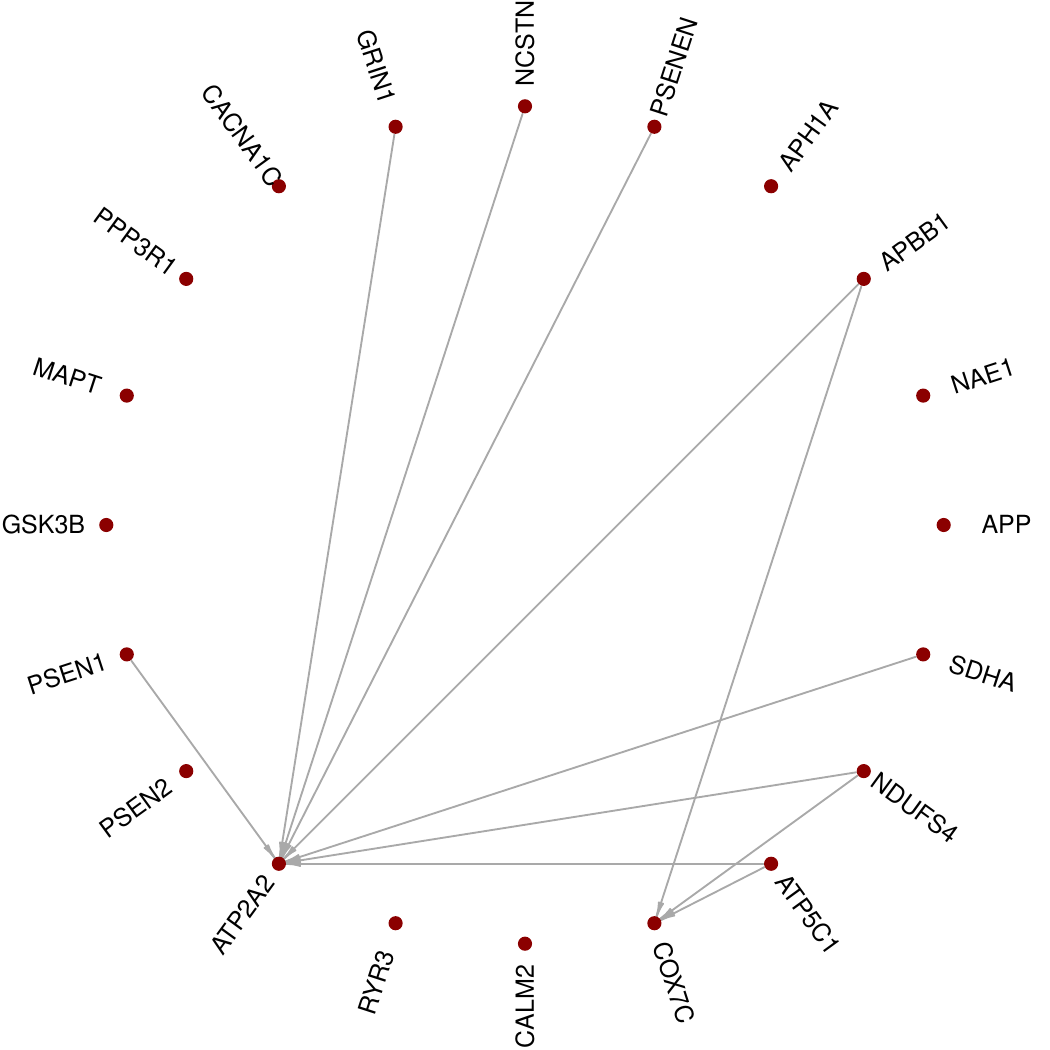}\label{fig:real-data-cn}}
  \caption{Reconstructed directed acyclic graphs for (a) AD-MCI and (b) 
CN groups.}
  \label{fig:real-data}
\end{figure}

As displayed in Figure \ref{fig:real-data}, the reconstructed DAGs exhibit some common and distinctive characteristics for the AD-MCI and CN groups. 
In the AD-MCI group, (1) directed edges $\text{GRIN1} \to \text{MAPT}$ and $\text{PSEN1}\to \text{GSK3B}$ agree with the reference pathways of the tau protein; (2) genes $\{ \text{APH1A, PSENEN, NCSTN, PPP3R1, APBB1, APP} \}$ have more directed connections, 
corresponding to the amyloid precursor protein. So do genes
$\{ \text{PSEN1, GSK3B} \}$ for the presenilin proteins.
By comparison, the genes participating in the amyloid precursor protein
and tau protein metabolism have fewer connections in the CN group \citep{o2011amyloid,palmqvist2020discriminative}.  
This observation seems consistent with previous studies that 
both genes may be involved in AD. 
Moreover, there are six and two non-root genes, respectively for the AD-MCI and CN groups. 

\begin{figure}
  \centering
  \subfloat[AD-MCI]{\includegraphics[width=0.4\textwidth]{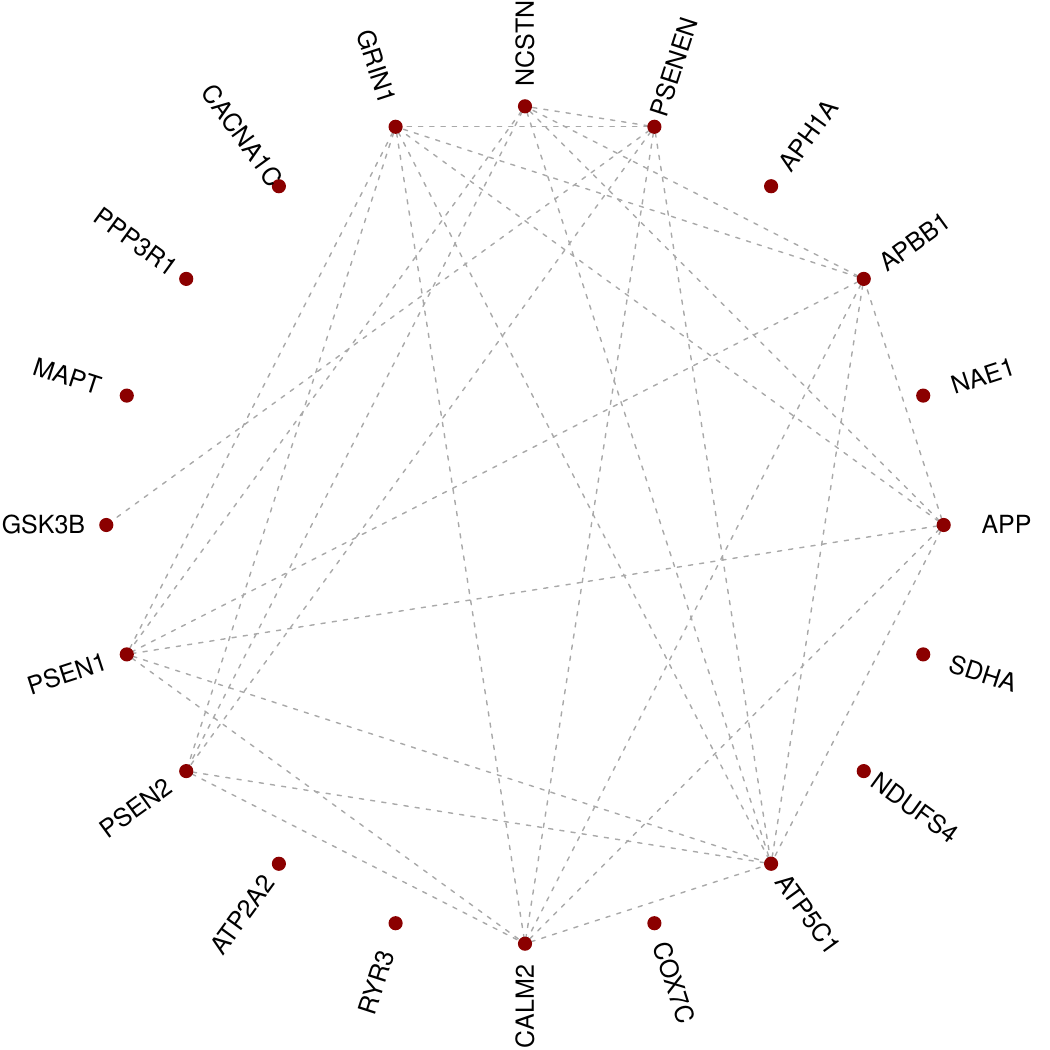}\label{fig:residual-ad}}
  \qquad
  \subfloat[CN]{\includegraphics[width=0.4\textwidth]{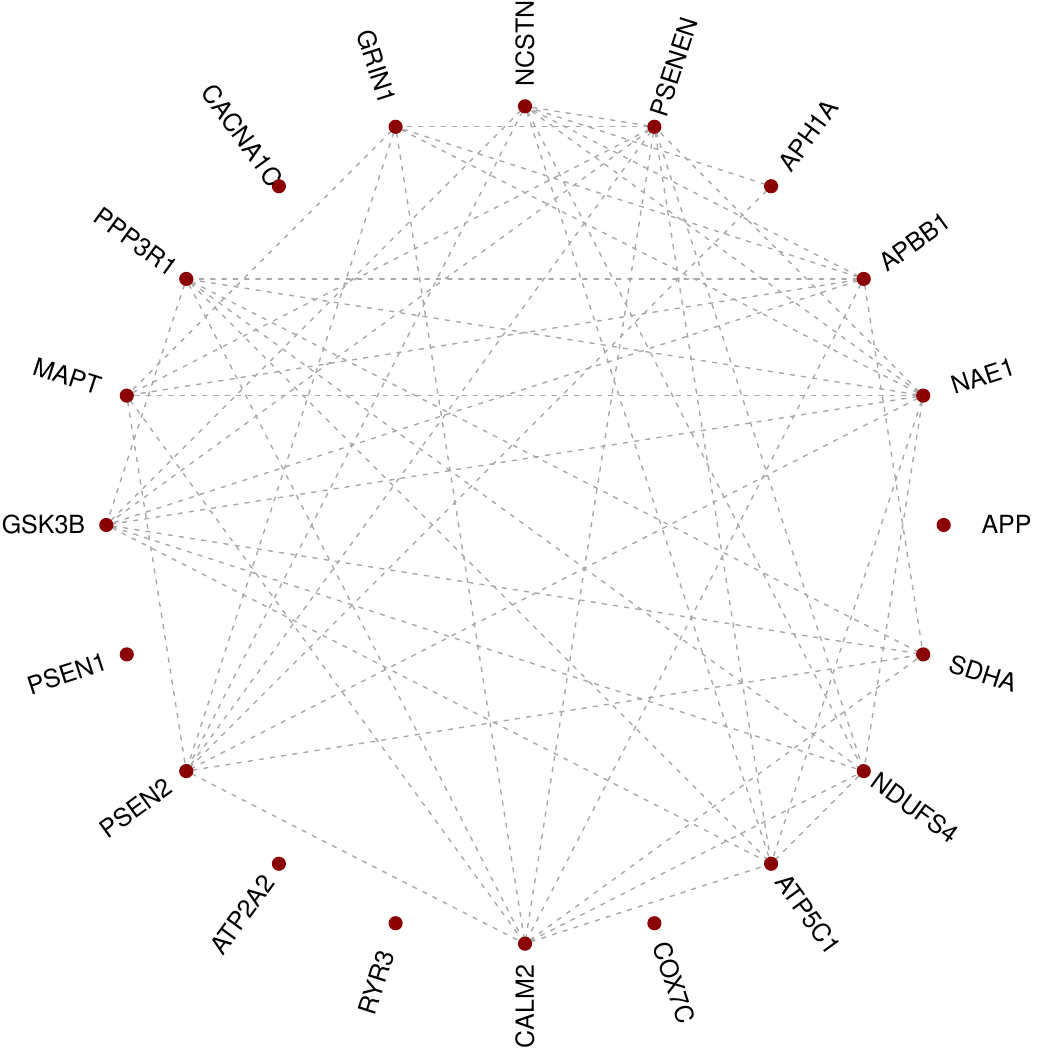}\label{fig:residual-cn}}
  \caption{Undirected graph displaying the estimated residual correlations of 
  $\widehat{\varepsilon} = \big(Y_{j} - \widehat{f}_j(Y_{\widehat \pa(j)})\big)_{j\in V}$, 
where a connection between two genes indicates the absolute value of residual correlation exceeds $0.15$. 
Edge connections from one gene to other multiple genes suggest the presence of confounders 
or nonzero off-diagonal elements of the covariance matrix $\Sigma$.}
  \label{fig:residual}
\end{figure}

For model diagnostics, we check the nonlinearity assumption on the gene expression levels. To this end, we compare a linear and a quadratic regression model for each non-root gene in the AD-MCI and CN groups in terms of their AIC values \citep{akaike1992information}. These models are fitted on the estimated parents of DeFuSE, and the quadratic model includes additional quadratic terms $(Y^2_{k})_{k\in\widehat{\pa}(j)}$ as covariates. For a linear or a quadratic model ${m}$ for a non-root variable $Y_j$, the AIC value is defined as 
\begin{equation}
\text{AIC}(\widehat{m}) = (n\widehat{\sigma}_{\text{FNN}}^2)^{-1}\sum_{i=1}^n (Y_{j}^{(i)} - \widehat{Y}_{j}^{(i)})^2 + 2 n^{-1}\dim(\widehat{m}),
\label{aic}
\end{equation}
where ${\widehat{m}}$ and $\widehat{\sigma}^2_{\text{FNN}}$ are the fitted model and the error variance estimated by FNN, $\widehat{Y}_{j}^{(i)}$ is the fitted values of $Y_{j}^{(i)}$, and $\dim(\widehat{m})$ denotes the number of parameters in model $\widehat{m}$. As suggested in Table \ref{table:model-checking-aic}, the quadratic model generally fits better than the corresponding linear model, as measured by AIC, suggesting that the nonlinearity assumption is approximately satisfied. Finally, the correlation plots of $\big(Y_{j}^{(i)} - \widehat{f}_j(Y^{(i)}_{\widehat \pa(j)})\big)_{j\in V}$; $i=1,\ldots,n$ in Figure \ref{fig:residual} exhibit the presence of (linear) hidden confounding as evident from the fact that many genes have multiple connections to other genes, indicating nonzero off-diagonals of $\Sigma$. This observation seems plausible due to the absence of some genes in the analysis.

\begin{table}[ht] 
  \footnotesize
  \centering
  \caption{The AIC values for quadratic and linear models fitted for each non-root gene, as defined in \eqref{aic}. A smaller AIC value indicates better model fitting.}
  \label{table:model-checking-aic}
  \begin{tabular}{l c c c c c c c c c}
  \hline
     Group       & \multicolumn{6}{c}{AD-MCI}                                & & \multicolumn{2}{c}{CN} \\
     \cline{2-7} \cline{9-10}
    Gene name    & APH1A  & PPP3R1 &  MAPT      & GSK3B  & COX7C   &  NDUFS4 & &  ATP2A2  & COX7C       \\ \hline
      Quadratic  & .717   & .656   &  .528     & .620   &  .356  & .606    & &  .572    & .304        \\
      Linear     & .701   & .732   &  .567     & .695   &  .395  & .657    & &  .656    & .349        \\ 
  \hline
  \end{tabular}
\end{table}

\section{Discussion}\label{sec:discussion}

This article proposes a novel method for learning functional causal relations with additive confounders. 
For modeling, we establish identifiability under {a sublinear growth condition} on the functional relationships. 
On this basis, we propose a novel method called DeFuSE and implement it with feedforward neural networks for scalability. 
Theoretically, we show that the proposed method consistently reconstructs all nonlinear causal relations.

One central message is that nonlinearity permits the separation of the nonlinear causal relationships from the confounding effects in model \eqref{model:confounding} with observational data only. 
As nonlinear causal discovery with hidden confounding remains understudied, we hope the work could inspire further research in this direction.

\appendix
\section{Appendix}

\subsection{Additional results on identifiability}

If $\Sigma\in\Psi$, the sublinear growth condition (Condition \ref{cond:sublinear}) is sufficient 
for identifying both $\{f_j\}_{1\leq j\leq p}$ and $\{\pa(j)\}_{1\leq j\leq p}$ in \eqref{model:confounding}. 
When this condition is not satisfied, it is still possible to establish identifiability under an alternative assumption. 
Now, we consider model \eqref{model:dependence} with additive functions,
\begin{equation}\label{equation:additive-model}
  Y_j = \sum_{k\in\pa(j)} f_{j,k}(Y_k) + \varepsilon_j, \quad j\in V = \{1,\ldots,p\},
\end{equation}
where $\{f_{j,k}\}$ are nonlinear and $\varepsilon \sim N(0,\Sigma)$.
Theorem \ref{thm:alternative-identifiability} establishes the identifiability of $\{\pa(j)\}_{1\leq j\leq p}$ in \eqref{equation:additive-model}, without the sublinear growth condition.

\begin{theorem}\label{thm:alternative-identifiability}
In \eqref{equation:additive-model}, assume that $Y_j-\E\left(Y_j\mid Y_{\vx(d)}\right)$ is not normally distributed for $\topdep_j>d$; $0 \leq d \leq d_{\max}$. For any univariate function $f$,
we define its equivalence class 
\begin{equation*}
   [f] = \{ \widetilde{f} : \widetilde{f}(z) = f(z) + \gamma z, \gamma\in\mathbb{R}  \}. 
\end{equation*}
If 
\begin{equation*}
  [f_{j,k}] \neq 
  \sum_{j'\in \vx(\topdep_j) }  \gamma_{j'} [f_{j',k}] \quad \text{ for all $\gamma_{j'}\in\mathbb{R}$; $j'\in\vx(\topdep_j)$, $j\in V = \{1,\ldots,p\}$,}
\end{equation*}
then $\{\pa(j)\}_{1\leq j\leq p}$ are uniquely identifiable.
\end{theorem}

The assumption that $Y_j-\E\left(Y_j\mid Y_{\vx(d)}\right)$ is not normal for $\topdep_j>d$ 
imposes constraints on the compositions of nonlinear functions,
which is automatically satisfied by sublinear growth functions when $\Sigma\in\Psi$ (Theorem \ref{thm:identifiability}).  
As suggested by the simulations in Section \ref{sec:numerical-example},
DeFuSE continues to perform well in recovering the DAG even when Condition \ref{cond:sublinear} and the additive function model \eqref{equation:additive-model} are both violated. 

\subsection{Regularity conditions}\label{sec:regularity-conditions}

We impose the following regularity conditions to establish the consistency of DeFuSE. 

\paragraph*{Metric entropy.} 
We define the bracketing $L_2$-metric entropy as a complexity measure 
of function spaces $\mathcal{G}_j = \left\{ g_j: g_j(\cdot,\star) 
= f_j\left(\cdot\right) + \left\langle \star, \beta_j \right\rangle \right\}$; $j=1,\ldots,p$, where $\cdot$ and $\star$ represent a $|\vx(\topdep_j)|$-dimensional vector, respectively. The bracketing $L_2$-metric entropy of $\mathcal{G}_j$ is the logarithm of the smallest $u$-bracket cardinality, $H(u,\mathcal{G}_j) = \log(\min\{ m : \mathcal S(u,m) \})$, where a $u$-bracket $\mathcal{S}(u, m)=\{g_1^-,g_1^+,\ldots,g_m^-,g_m^+\}\subseteq L_2(P)$ is a set of functions 
such that (i) $\max_{1\leq k\leq m}\|g_k^- - g_k^+\|_{L_2}\leq u$ and (ii) for any $g \in \mathcal{G}_j$ there exists $g_k^-\leq g \leq g^+_k$ almost surely.

\begin{condition} \label{cond:entropy}
For some positive $\epsilon_n<1/2$, 
\begin{equation*}
\max_{1\leq j\leq p}
\max_{\{ A: |A|\leq |\pa^\circ(j)| \}} 
\int^{\sqrt{2}\epsilon_n}_{\epsilon_n^2/256} H^{1/2}(u/c_1, 
\mathcal{G}_j(A)) du
\leq c_2 \sqrt{n}\epsilon_n^2,
\end{equation*}
where $\mathcal{G}_j(A) = \left\{ g_j \in \mathcal F_j : A = \ARG(f_j),\ \|g_j-g^\circ_j\|_{2}\leq 2\epsilon_n \right\}$ is the $2\epsilon_n$-neighborhood of $g^\circ_j$ on the index set of effective arguments $A$.
\end{condition}

In view of Condition \ref{cond:entropy}, the error rate $\epsilon_n$ is determined by solving the integral equation in $\epsilon_n$. Such a condition has been used to quantify the convergence rate of sieve estimates \citep{wong1995probability, van2000empirical}. The entropy results are available for many function classes, such as the FNN in Theorem \ref{thm:fnn}.

\paragraph*{Sparsity and confounding.}

Next, we impose a regularity condition on sparsity and confounding structures, requiring the true support of $g^\circ_j$, the maximum depth $d_{\max}$, and the error variance not to increase with the sample and graph sizes $(n,p)$. 

\begin{condition} \label{cond:confounding}
  Assume $\kappa^\circ=\max_{1\leq j\leq p}|{\pa}^\circ(j)|$, $\varsigma^\circ = \max_{1\leq j\leq p}\|\beta^\circ_j\|_0$,
  $d_{\max} = \max_{1\leq j\leq p} \topdep_j$, and $c_{-}\leq\lambda_{\min}(\Sigma)\leq \lambda_{\max}(\Sigma)\leq c_+$ 
  are independent of $(p,n)$, 
  where $\lambda_{\min}(\Sigma)$ and $\lambda_{\max}(\Sigma)$ are the smallest and largest eigenvalues of $\Sigma\in\Psi$.
\end{condition}

\subsection{Implementation details}\label{sec:implementation}

The code is open-sourced at \url{https://github.com/chunlinli/defuse}.

\paragraph*{Training and hyperparameter tuning for DeFuSE.}

Training and tuning a neural network requires intensive computation. Following the conventional practice of deep learning, we split the original sample into training and validation sets with a partition ratio 9:1, and use on-the-fly evaluation over the validation set for tuning during the training process.

To tune hyperparameters $\kappa_j,\varsigma_j$ in \eqref{deep}, we adopt a greedy strategy combined with an asynchronous-synchronous training technique since it is unnecessary to identify the exact value of $\varsigma_j$, c.f., Theorem \ref{thm:variable-selction}. We first optimize \eqref{deep} in $\beta_j$ with $\theta_j = 0$, subject to the sparsity constraint $\sum_{k\in\vx(d)}\min(|\beta_{j,k}|/\tau,1)\leq \varsigma_j$, followed by selecting $\varsigma_j\in\{0,1,\ldots,|\vx(d)|\}$ that minimizes the mean squared error on the validation set. Throughout, we fix $\tau = 0.05$ as a signal-noise threshold. This stage intends to perform a sparsity-constrained linear regression, so it is very efficient in computing. Next, given the selected variable set 
$B = \{ k: |\beta_{jk}|\geq \tau \}$ in \eqref{deep}, we estimate $(\theta_j,\beta_{j,B})$ with $\beta_{j,B^c} = 0$ by minimizing 
\begin{equation*}
\begin{split}
\min_{\theta_j} \quad \
\sum_{i=1}^n \Big(Y_{j}^{(i)} - f_j\big(Y_{\vx(d)}^{(i)}\big) - \big\langle \widehat{\xi}_{\vx(d)}^{(i)}, \beta_{j,B}\big\rangle \Big)^2, \quad\text{s.t.} \ \ \quad
\sum_{k\in \vx(d)}
\min(\|W^{1}_k \|/\tau,1)\leq \kappa_j.
\end{split}
\end{equation*}
To leverage the automatic differentiation in modern deep learning libraries, we consider its regularized version with $\kappa_j$ replaced by a hyperparameter
$\lambda_j > 0$: 
\begin{equation*}
  \begin{split}
  \min_{\theta_j} \quad& \ 
  \sum_{i=1}^n \Big(Y_{j}^{(i)} - f_j\big(Y_{\vx(d)}^{(i)}\big) - \big\langle \widehat{\xi}_{\vx(d)}^{(i)}, \beta_{j,B}\big\rangle \Big)^2 + \lambda_j \sum_{k\in \vx(d)}
  \min(\|W^{1}_k \|/\tau,1).
  \end{split}
\end{equation*}
where $\lambda_j > 0$ controls the degree of regularization. Then, after the regularized optimization is completed, we tune $\kappa_j \in \{0,1,\ldots,|\vx(d)|\}$ using the top $\kappa_j$ variables (sorted by weight $\|W_k^1\|$) among all variables and masking the rest.
To speed up the computation, we also implement a nonparametric screening procedure \citep{azadkia2021simple} for variable selection.

In our experiments, we use an adaptive regularization approach for $\lambda_j > 0$ during training, similar to adaptive learning rate scheduling. Specifically, we consider three candidate values 
$\lambda_j\in\{0.0001, 0.001, 0.05\}$. The training process starts with $\lambda_j = 0.0001$ and gradually increases $\lambda$ to achieve better validation performance by inducing more 
sparsity. Based on our limited experience, this adaptive regularization strategy is effective and can be combined with other deep learning techniques such as early stopping. 

For network structure, we use an FNN with one hidden layer and 50 hidden neurons. 
For optimization, we use the Adam optimizer \citep{kingma2014adam} with a learning rate $0.1$ and various numbers of epochs $\{ 250, 500, \ldots, 4000 \}$ in our experiments. Then we choose the best-performing model.

\paragraph*{Other methods.}

The R packages \texttt{CAM}, \texttt{pcalg}, and \texttt{lrpsadmm} are available at \url{https://github.com/cran/CAM}, \url{https://github.com/cran/pcalg}, and \url{https://github.com/benjaminfrot/lrpsadmm}, respectively.
The Python program \texttt{notears} is available at \url{https://github.com/xunzheng/notears}.
We use their default settings for CAM, NPTEARS, LRpS-GES, and RFCI.

\bibliographystyle{apalike}
\bibliography{ref}

\newpage

\begin{center}
    \Large\bf Supplementary Materials for ``Nonlinear causal discovery with confounders''
\end{center}

\section*{Technical proofs}

In what follows, $c_j$'s and $C$ denote generic constants. 

\begin{lemma}\label{lemma:gaussian-tail}
  If $W \sim N(0,1)$ and $t>0$, then 
  \begin{equation*}
    \sqrt{\frac{2}{\pi}} \frac{t}{1 + t^2} e^{-t^2/2} \leq P(|W|\geq t) \leq \sqrt{\frac{2}{\pi}} \frac{1}{t}e^{-t^2/2}.
  \end{equation*}
\end{lemma}

By Lemma \ref{lemma:gaussian-tail}, if $W\sim N(0,\sigma_W^2)$, $\sigma_W^2 = \inf \Big\{  c : \lim_{t\to\infty} P(|W|>t) \exp(t^2/2c) = 0   \Big\}$.

\begin{lemma}\label{lemma:sublinear-gaussian}
Assume $X = (X_1,\ldots,X_q) \sim N(0,\Sigma_X)$ and 
  $\sigma_\gamma^2 = \gamma^\top\Sigma_X\gamma$.
  If $Z = f(X) + \langle X, \gamma\rangle$, then under Condition 1, 
  \begin{equation*}
    \sigma_\gamma^2 = \inf \Big\{  c : \lim_{t\to\infty} P(|Z|>t) \exp(t^2/2c) = 0   \Big\}.
  \end{equation*}  
\end{lemma}
\begin{proof}
  Note that $P(|Z|>t) = P(|f(X) + \gamma^\top X| > t ) = P( |\gamma^\top X|   > t / |1 + f(X)/\gamma^\top X| )$.
  On event $\{ |\gamma^\top X| > t / |1 + f(X)/\gamma^\top X| \}$, when $t\to\infty$, we have $|\gamma^\top X|\to\infty$.
  Hence, by Condition 1, for any small $\varepsilon > 0$, when $t$ is large enough, 
  \begin{equation*}
    P(|\gamma^\top X| > t/(1 - \varepsilon) ) \leq   P(|Z|>t) \leq  P(|\gamma^\top X| > t/(1 + \varepsilon) ).
  \end{equation*}
  Let $W = \gamma^\top X/\sigma_\gamma$. 
  By Lemma \ref{lemma:gaussian-tail},
  \begin{equation*}
    \begin{split}
      P(|W|> t/\sigma_\gamma(1+\varepsilon)) 
      &\leq \sqrt{\frac{2}{\pi}} \frac{\sigma_\gamma(1+\varepsilon)}{t} e^{-t^2/2\sigma_\gamma^2(1+\varepsilon)^2},\\
      P(|W|>t/\sigma_\gamma(1-\varepsilon))
      &\geq \sqrt{\frac{2}{\pi}} \frac{t \sigma_\gamma (1-\varepsilon)}{ \sigma_{\gamma}^2(1-\varepsilon)^2 + t^2}
      e^{-t^2/2\sigma_\gamma^2(1-\varepsilon)^2}.
    \end{split}
  \end{equation*} 
  As a result, $\inf \Big\{  c : \lim_{t\to\infty} P(|Z|>t) \exp(t^2/2c) = 0   \Big\} = \sigma_\gamma^2$.
\end{proof}

For identifiability, we first prove Theorem 2, followed by Theorem 1.

\begin{proof}[Proof of Theorem 2]
  Note that $\E(\varepsilon_j\mid Y_{\vx(d)})=\E(\varepsilon_j \mid \xi_{\vx(d)})$ for any $d_j = d$.
  By (2), we have 
  \begin{equation*}
      \E(Y_j \mid Y_{\vx(d)})=f_j(Y_{\pa(j)})+ \E(\varepsilon_j \mid \xi_{\vx(d)}).
  \end{equation*}
  Transforming $\varepsilon =(\varepsilon_{V(1)},\varepsilon_{V(2)\setminus V(1)},\ldots,\varepsilon_{V\setminus V(d_{\max})})$ to $\xi=(\xi_{V(1)},\xi_{V(2)\setminus V(1)}\ldots,\xi_{V\setminus V(d_{\max})})$ can be regarded as a block Gram-Schmidt process, where $\xi_{V(1)},\xi_{V(2)\setminus V(1)}\ldots,\xi_{V\setminus V(d_{\max})}$ are uncorrelated.
  Thus, $(\varepsilon_j, \xi_{\vx(d)})$ follows
  a joint Gaussian distribution.
  The desired result follows from the fact that 
  $\E(\varepsilon_j \mid \xi_{\vx(d)}) = \langle \xi_{\vx(d)}, \beta_j\rangle$. This completes the proof.
\end{proof}

\begin{lemma}\label{lemma:zero-measure}
Under Condition 1, 
the set $\Psi^c$ is closed and nowhere dense in $\{ \Sigma: \Sigma \succ 0 \}$.
Moreover, $\Psi^c$ has zero Lebesgue measure.
\end{lemma}
\begin{proof}
  Note that $\Sigma$ can be reparameterized by 
  \begin{equation*}
      \Big\{\big(\Var(\xi_{\vx(d)\setminus \vx(d-1) }), \beta_j\big): j\in\vx(d),\ d=1,\ldots,d_{\max}+1\Big\}.
  \end{equation*}
  Moreover, $f_j(Y_{\pa(j)})$ can be written as a function of $\xi_{\vx(d)}$, 
that is, $\widetilde{f}_j(\xi_{\vx(\topdep_j)})$. 
  Then 
  \begin{equation*}
    Y_j = f_j(Y_{\pa(j)}) + \langle \xi_{\vx(\topdep_j)}, \beta_j\rangle + \xi_j = \widetilde{f}_j(\xi_{\vx(\topdep_j)}) + \langle \xi_{\vx(\topdep_j)}, \beta_j\rangle + \xi_j.
  \end{equation*}
  Let $\topdep_j > d$. Suppose $Y_j \mid Y_{\vx(d)}$ is normal with mean $\E(Y_j \mid Y_{V(d)})$ and constant variance. 
  Note that the distribution $Y_j \mid Y_{\vx(d)} = y$ is the same as $Y_j\mid \xi_{\vx(d)} = x$ for some $x$. 
  Fixing $\xi_{\vx(d)} = x$, we have 
  \begin{equation*}
    Y_j \mid \{Y_{\vx(d)} = y\} = \underbrace{\widetilde{f}_j\big(x, \xi_{\vx(\topdep_j)\setminus \vx(d) }\big) + \big\langle \xi_{\vx(\topdep_j)\setminus\vx(d)},{\beta}_{j,\vx(\topdep_j)\setminus\vx(d)} \big \rangle}_{Z} + \big\langle x, \beta_{j,\vx(d)} \big\rangle + \xi_j. 
  \end{equation*}
By Lemma \ref{lemma:sublinear-gaussian}, $Z$ has to be normal with constant variance $\widetilde{\sigma}^2 = \Var(\langle \xi_{\vx(\topdep_j)\setminus\vx(d)},{\beta}_{j,\vx(\topdep_j)\setminus\vx(d)}  \rangle)$. 
  For simplicity, denote $h(\xi_{\vx(\topdep_j)\setminus \vx(d) }) = \widetilde{f}_j(x, \xi_{\vx(\topdep_j)\setminus \vx(d) })$,
  $\zeta = \xi_{\vx(\topdep_j)\setminus \vx(d) }$ and $\gamma = {\beta}_{j,\vx(\topdep_j)\setminus\vx(d)}$.
  For each $x$, this implies an infinite set of moment conditions for $Z$,  
  \begin{equation*}
    \begin{split}
      \E h^2 + 2 (\E h  \zeta)^\top \gamma  &= 0,\\
      \E h^3 + 3 (\E h^2\zeta)^\top \gamma + 3 \gamma^\top (\E h\zeta\zeta^\top) \gamma &= 0,\\
      &\cdots\ .
    \end{split}
  \end{equation*}
  The solution(s) $\gamma$ to the above algebraic equations forms a closed, measure zero, and nowhere dense set in $\mathbb{R}^{|\vx(\topdep_j)\setminus \vx(d)|}$.
  For $\topdep_j > d$, the corresponding $\beta_j$ depends on $\Var(\xi_{\vx(d')\setminus \vx(d'-1)})$ and $\beta_{j'}$ for $j'\in\vx(d')$; $d'\leq d$. 
  Thus, $\Psi^c$ is a finite union of closed and measure zero sets, and as a result, it is nowhere dense. This completes the proof.
\end{proof}

\begin{proof}[Proof of Theorem 1]

Based on Lemma \ref{lemma:zero-measure}, we identify $\vx(1) \subseteq \cdots \subseteq \vx({d_{\max}}+1)=V$ of the true 
graph. It remains to show $\{f_k\}_{1\leq k\leq p}$ are identifiable.
Now, suppose $\xi_{\vx(d)}$ are given. 
By Theorem 2, $\E(Y_j\mid Y_{\vx(d)}) =
f_j(Y_{\pa(j)}) + \left\langle \xi_{\vx(d)}, \beta_j\right\rangle$.
If there exist $(\tilde{f}_j,\tilde{\beta}_j,{\tilde{\pa}(j)})$ such that
$\E(Y_j\mid Y_{\vx(d)})
=\tilde{f}_j(Y_{\tilde{\pa}(j)}) + \langle \xi_{\vx(d)}, \tilde\beta_j\rangle$.
Then,
\begin{equation} \label{identification}
f_j(Y_{\pa(j)}) - \tilde{f}_j(Y_{\tilde{\pa}(j)}) 
= \langle \xi_{\vx(d)}, \tilde{\beta}_j - \beta_j\rangle.
\end{equation}
To prove that $f_j(Y_{\pa(j)}) - \tilde{f}_j(Y_{\tilde{\pa}(j)}) = 0$ almost surely, we first show that $f_j(Y_{\pa(j)}) - \tilde{f}_j(Y_{\tilde{\pa}(j)})$ is constant.
Otherwise, $F(Y_A) = f_j(Y_{\pa(j)}) - \tilde{f}_j(Y_{\tilde{\pa}(j)})$ 
functionally depends on $Y_A$ for a nonempty subset $A\subseteq \pa(j)\cup\tilde{\pa}(j)$, which we assume, without loss of generality, that 
$A = \ARG(F)$ is minimal in that $F(Y_{A})$ depends on all variables indexed by $A$.
Consider any $k\in A$ with $\topdep_k = \max_{l\in A} \topdep_l$.
Since $\xi_k=Y_k- \E(Y_k \mid Y_{\vx(\topdep_k)})$, we must have that $k\in B=\{ l\in \vx({d}) : \beta_{jl}\neq \tilde{\beta}_{jl} \}$ and $\topdep_k = \max_{l\in B}\topdep_l$.
Moreover, the only term involving $Y_k$ in the right-hand side of \eqref{identification} is
a term $(\tilde{\beta}_{jk}-\beta_{jk}) Y_k$,
because by definition $Y_k$ does not appear in any $\xi_{l}$ for any $l\in B$ such that $l\neq k$.
If $\tilde{\beta}_{jk}-\beta_{jk}\neq 0$, 
then the right-hand side of \eqref{identification} 
becomes $\sum_{l\in B}(\tilde{\beta}_{jl} - \beta_{jl}) \xi_{l} \mid Y_{\vx({\topdep_k})}$, 
which is Gaussian.
However, on the left-hand side, 
$F(Y_A) \mid Y_{ \vx({\topdep_k}) } = (f(Y_{\pa(j)}) - \tilde{f}(Y_{\tilde{\pa}(j)})) 
\mid Y_{ \vx({\topdep_k}) }$ is not Gaussian under Condition 1 by Lemma \ref{lemma:gaussian-tail}, which leads to a contradiction.
So $A=\emptyset$ and $F(Y_A)$ is constant. Note that the right-hand side of \eqref{identification} has mean zero, which completes the proof.
\end{proof}

\begin{lemma}\label{lemma:variable-selection}
Assume that Conditions 2-3, 5-6 are met. 
Let $\widehat{g}_j$ be an $\delta_n^2$-minimizer of a least squares regression criterion such that 
  \begin{equation*}
    \left\| \bm Y_{j} - \widehat g_j\left( \bm Y_{\vx(\topdep_j)},\bm \xi_{\vx(\topdep_j)}\right)\right\|^2
    \leq \min_{g_j\in\mathcal F_j} 
    \left\|\bm Y_{j} - g_j\left(\bm Y_{\vx(\topdep_j)},\bm \xi_{\vx(\topdep_j)}\right)\right\|^2 
    + c_8n\delta_n^2,
  \end{equation*}
with $\delta_n^2\geq \epsilon_n^2$. Then
\begin{equation*}
P\big(\widehat{\pa}(j)\neq \pa^\circ(j)\big)\leq c_7 \exp(-c_5 n \delta_n^2 - 
\log n ); \quad j=1,\ldots,p. 
\end{equation*}
\end{lemma}

\begin{proof}[Proof of Lemma \ref{lemma:variable-selection}]
  By Condition 3, any $f_j\in\mathcal F_j$ with a wrong support set 
$\ARG(f_j)\neq \pa^\circ(j)$ satisfies 
$\|g_j - g_j^\circ\|^2_2\geq 4c_3\epsilon_n^2$.
However, by Condition 2, $\|g^*_j-g^\circ_j\|_{L_2}
\leq c_3\epsilon_n^2 <4c_3\epsilon_n^2$, implying
that $f^*_j$ has the same support of $f^\circ_j$ or $\ARG(f^*_j)=\pa^\circ(j)$. 
Let $d = \topdep_j$.

\textit{Step 1. Partitioning.} 
Given a class $\{ A : A\neq \pa^\circ(j), |A|\leq |\pa^\circ(j)| \}$ of candidate augmented sets of $\ARG(g_j)$,
we partition $A$ as  
$A=(A\setminus \pa^\circ(j))\cup(A\cap \pa^\circ(j))$; $j=1,\ldots,p$. 
Now consider a partition of $\mathcal{F}_j$. Let 
\begin{equation*}
    \mathcal{E}(\nu_1,\nu_2) =\left\{ g_j\in \mathcal{F}_j : 
    \begin{aligned}
\ARG(f_j) \neq \pa^\circ(j), \ |A\cap \pa^\circ(j)|=\nu_1, \ |A\setminus \pa^\circ(j)|=\nu_2, \\
(|\pa^\circ(j)| - \nu_1)D_{\min} \leq \|g - g_j^*\|_{L_2}^2
    \end{aligned}
 \right\}
\end{equation*}
be a subclass of functions of $\mathcal{F}_j$; 
$\nu_1=0,\ldots,|\pa^\circ(j)|-1$ and $\nu_2 = 1,\ldots, |\pa^\circ(j)|-\nu_1$.
Then functions in $\mathcal{E}(\nu_1,\nu_2)$ have at most 
$\binom{|\pa^\circ(j)|}{\nu_1}\binom{p-|\pa^\circ(j)|}{\nu_2}$ different supports. 
By definition, 
\begin{equation*}
    \Big\{g_j\in \mathcal{F}_j, A = \ARG(f_j) : A\neq \pa^\circ(j), |A|\leq |\pa^\circ(j)| \Big\} 
\subseteq \bigcup_{\nu_1=0}^{|\pa^\circ(j)|-1}\bigcup_{\nu_2=1}^{|\pa^\circ(j)|}\mathcal{E}(\nu_1,\nu_2).
\end{equation*}
Denote by the log-likelihood $L_{j}(g_j) = -\|\bm Y_{j}-g_j\left(\bm Y_{\vx(\topdep_j)}, 
\bm \xi_{\vx(\topdep_j)} \right) \|^2/2 \sigma^2_j=1$. Here, without loss of generality, we assume that
$\sigma^2_j=1$ and $|\pa^\circ(j)| \geq 1$. Using the previously established fact that
$\ARG(g^*_j)=\pa^\circ(j)$, we have
\begin{equation*}
\begin{split}
P(\widehat{\pa}(j)\neq \pa^\circ(j)) =\
& P^*\left( \sup_{\{g_j\in \mathcal F_j: \ARG(f_j)\neq \pa^\circ(j), |\ARG(f_j)|\leq |\pa^\circ(j)| \} }
(L_{j}(g_j) - L_{j}(g_j^*)) \geq -c_8 n \epsilon_n^2 \right)\\
\leq\ &\sum_{\nu_1=0}^{|\pa^\circ(j)|-1}\sum_{\nu_2=1}^{|\pa^\circ(j)|-\nu_1} 
P^*\left( \sup_{g \in \mathcal{E}(\nu_1,\nu_2)} (L_{j}(g_j) - L_{j}(g_j^*)) \geq -c_8 n \epsilon_n^2 \right),\\
\end{split}
\end{equation*}
where $P^*$ denotes the outer probability. 

\textit{Step 2. Large-deviation bounds.} 
Let $\Delta_n=\max_{1\leq j\leq p} \Big(\E(p_{g^\circ_j}/p_{g_j^*}) - 1\Big)$ be the Kullback-Leibler 
divergence, where $p_{g_j}=p_{g_j}(Y_j,Y_{\vx(d)},\xi_{\vx(d)})$ is the joint probability density function for $(Y_j,Y_{\vx(d)},\xi_{\vx(d)})$. 
By (2), $p_{g_j}(Y_j,Y_{\vx(d)},\xi_{\vx(d)}) = \exp(-(Y_j - g_j(Y_{\vx(d)},\xi_{\vx(d)}))^2/2)$. 
By Condition 2, $\E(p_{g^\circ_j}/p_{g_j^*})$ equals to 
\begin{equation*}
  \begin{split}
    & \E \exp\left(  \xi_j (g^*_j(Y_{\vx(d)},\xi_{\vx(d)}) - g^\circ_j(Y_{\vx(d)},\xi_{\vx(d)}))  
    + \frac{(g^*_j(Y_{\vx(d)},\xi_{\vx(d)}) - g^\circ_j(Y_{\vx(d)},\xi_{\vx(d)}))^2}{2}  \right)\\
    =\ & \E \exp\left(  |g^*_j(Y_{\vx(d)},\xi_{\vx(d)}) - g^\circ_j(Y_{\vx(d)},\xi_{\vx(d)})|^2 \right) \\
    =\ & \E \exp( |f^*_j(Y_{\vx(d)}) - f^\circ_j(Y_{\vx(d)})|^2 )\\
    \leq\ & c \| f_j^* - f_j^\circ\|_{L_2}^2 + 1 \\
    =\ & c \|g^*_j - g^\circ_j\|_{L_2}^2 + 1.
  \end{split}
\end{equation*}
Note that, for some constant $c>0$, 
$c\|g_j - g^*_j\|_{L_2}^2 \leq h^2(g_j,g_j^*) \equiv 1 - \exp(-\| g_j - g^*_j\|^2_2/8)$
when $\|g_j - g^*_j\|_{L_2}^2\leq 1$, where $h$ is the Hellinger-distance.
By Theorem 3 of \citet{wong1995probability} with $\delta_n(1)=\Delta_n$ there, 
under Conditions 3 and 5, 
there exists a constant $c_5 > 0$ such that 
\begin{equation*}
\begin{split}
&P^*\left( \sup_{g_j \in \mathcal{E}(\nu_1,\nu_2)} L_{j}(g_j) - L_{j}(g_j^*) \geq -c_8 n \epsilon_n^2 \right)\\
\leq& 5 {\binom{p-|\pa^\circ(j)|}{\nu_1} \binom{|\pa^\circ(j)|}{\nu_2}} 
\exp( -c_5 n (|\pa^\circ(j)|-\nu_1) D_{\min} + n(\Delta_n - 1)).
\end{split}
\end{equation*}
Thus, $P(\widehat{\pa}(j)\neq \pa^\circ(j))$ is upper bounded by 
\begin{equation*}
\begin{split}
&\sum_{\nu_1=0}^{|\pa^\circ(j)|-1}\sum_{\nu_2=1}^{|\pa^\circ(j)|-\nu_1} 
5 \binom{|\pa^\circ(j)|}{\nu_1} \binom{p-|\pa^\circ(j)|}{\nu_2} 
\exp( -c_5 n (|\pa^\circ(j)|-\nu_1) D_{\min} + n(\Delta_n - 1))\\
\leq & \sum_{\nu_1=0}^{|\pa^\circ(j)|-1} 5 \binom{|\pa^\circ(j)|}{\nu_1} 
\exp(- (|\pa^\circ(j)|-\nu_1) ( c_5 n D_{\min} - \log p) + n(\Delta_n - 1) ) \\
\leq & c_7 \exp(-c_5n D_{\min} +\log p + n(\Delta_n - 1)).
\end{split}
\end{equation*}
This completes the proof. 
\end{proof}

\begin{lemma}\label{lemma:empirical-norm}
Under the assumptions of Theorem 3,
\begin{equation*}
\begin{split}
   & P\Big(\|\widehat{g}_j(\bm Y_{\vx(d)},\bm \xi_{\vx(d)}) - g^\circ_j(\bm Y_{\vx(d)},\bm \xi_{\vx(d)})\|^2 \geq  c_8 n \epsilon_n^2 \Big) \\
 \leq\ & 3\exp( - (1-c_6) n \epsilon^2_n ) + c_7\exp(-c_5n\epsilon_n^2 - \log n),
\end{split}
\end{equation*}
provided that $P(\widehat{\pa}(j)\neq \pa^\circ(j))\leq c_7\exp(-c_5n\epsilon_n^2 - \log n)$; $j=1,\ldots,p$.
\end{lemma}
\begin{proof}[Proof of Lemma \ref{lemma:empirical-norm}]
From (4), $\|\bm Y_j - \widehat{g}_j(\bm Y_{\vx(d)},\bm \xi_{\vx(d)})\|^2\leq \|\bm Y_j-g^\circ(\bm Y_{\vx(d)},\bm \xi_{\vx(d)})\|^2 + 4n\epsilon_n^2$. 
Some simple algebra yields that  
  \begin{equation}
\label{empirical}
    \begin{split}
     \Big\|\widehat{g}_j(\bm Y_{\vx(d)},\bm \xi_{\vx(d)}) - g^\circ_j(\bm Y_{\vx(d)},\bm \xi_{\vx(d)})\Big\|^2
      \leq 2 \sum_{i=1}^n \xi^{(i)}_{j} 
      \left(\widehat{g}_j(Y_{\vx(d)}^{(i)},\xi_{\vx(d)}^{(i)}) - g_j^\circ(Y_{\vx(d)}^{(i)},\xi_{\vx(d)}^{(i)})\right) + 4 n \varepsilon_n^2,
    \end{split}
  \end{equation}
where $\xi_{j}^{(i)}=Y_{j}^{(i)}- g^\circ\left(Y_{\vx(d)}^{(i)},\xi_{\vx(d)}^{(i)}\right) 
= Y_{j}^{(i)}- \E\left(Y_{j}^{(i)}\mid Y_{\vx(d)}^{(i)},\xi_{\vx(d)}^{(i)}\right)$ 
has mean zero and is independent of $(Y_{\vx(d)}^{(i)},\xi_{\vx(d)}^{(i)})$.

  Next, we apply Theorem 3 of \citet{shen1994convergence} to bound the empirical process 
  \begin{equation*}
    \begin{split}
      n^{-1} \sup_{\{\|g_j - g^\circ_j\|^2_2\leq C\epsilon_n^2\}} \sum_{i=1}^n 
      \xi_{j}^{(i)}\left({g}_j(Y_{\vx(d)}^{(i)},\xi_{\vx(d)}^{(i)}) - g_j^\circ(Y_{\vx(d)}^{(i)},\xi_{\vx(d)}^{(i)})\right).
    \end{split}
  \end{equation*}
To verify the conditions there, we assume, without loss of generality, 
that $\Var(\xi_j) = 1$ subsequently.
It suffices to consider $\{ \widehat{\pa}(j) = \pa(j) \}$. 
  Define the function space 
  \begin{equation*}
      \mathcal H = \Big\{ h : h(Y_{\vx(d)},\xi_{\vx(d)},\xi_j) 
  = \xi_{j} (g_j(Y_{\vx(d)},\xi_{\vx(d)}) -  g^\circ_j(Y_{\vx(d)},\xi_{\vx(d)}) ) \Big \}.
  \end{equation*}
  Then $\|h\|_{L_2} = \|g_j - g_j^\circ\|_{L_2}$. 
  Note that $\sup_{\mathcal H} |h|\leq T = C \sqrt{\log n + \log p} \times \sqrt{\log n}$ almost surely in $P$,
  $\E h = 0$, $\sup_{\mathcal H} \Var(h) \leq C\epsilon_n^2$. 
  Let $v = C\epsilon_n^2 T$, $M = n^{1/2}v/8T$. 
  By Condition 5, 
  \begin{equation*}
    \frac{\epsilon_n}{\sqrt{T}} H_B(\epsilon_n/\sqrt{T}, \mathcal H) 
    = \frac{\epsilon_n}{\sqrt{T}} H_B(\epsilon_n/\sqrt{T},\mathcal F_j^A)\leq \frac{\sqrt{n} \epsilon_n^2 / T}{70}. 
  \end{equation*}
  Moreover, $\epsilon_n\leq T$ and by Condition 5,  
  \begin{equation*}
    \int_{v/64T}^{v^{1/2}} H_B^{1/2}(u,\mathcal H) du \leq \frac{ \sqrt{n} \epsilon_n^2 }{2^{13}}.
  \end{equation*}
  By Theorem 3 of \citet{shen1994convergence}, we have 
  \begin{equation*}
    P\Big( n^{-1} \sup_{\{\|g_j - g^\circ_j\|^2_2\leq C\epsilon_n^2\}} 
    |\sum_{i=1}^n \xi_{j}^{(i)}({g}_j(Y_{\vx(d)}^{(i)},\xi_{\vx(d)}^{(i)}) - g_j^\circ(Y_{\vx(d)}^{(i)},\xi_{\vx(d)}^{(i)}))|
    \geq \epsilon_n^2 \Big) \leq 3\exp( - (1-c_6) n \epsilon^2_n ).
  \end{equation*}
The desired result follows immediately. 
\end{proof}

\begin{proof}[Proof of Theorem 3]
 
We prove Theorem 3 by induction for $\vx(1),\ldots,\vx({d_{\max}})$. 
First, note that no estimation is needed for $\vx(1)$. 
For $j\in V\setminus \vx(1)$, we bound $P(\widehat{\pa}(j)\neq \pa^\circ(j))$
as well as $\|\widehat{g}_j-g_j^\circ\|_{L_2}$. The proof proceeds in two steps.

\textit{Step 1. Bounds for error-in-variable $\bm \xi_{\vx(d)}$.} 
For $j\in \vx(d)$ with $d=\topdep_j\geq 1$, 
let $\widehat{g}_j$ be the estimated function via (7) based on error-in-variables 
$(\bm Y_{\vx(d)}, \widehat{\bm \xi}_{\vx(d)})\in\mathbb{R}^{n\times2|\vx(d)|}$, 
where $\widehat{\bm \xi}_{\vx(d)}=(\widehat{\bm \xi}_k)_{k \in \vx(d)}$ 
and $\widehat{\bm \xi}_k = \bm Y_k - \widehat{g}_k(\bm Y_{\vx(d)},\widehat{\bm \xi}_{\vx(d)})$. Let
$\bm \xi_k=\bm Y_k - g^\circ_k(\bm Y_{\vx(d)}, \bm \xi_{\vx(d)})$ be the oracle residual vector. 
We bound $\|\bm \xi_j-\widehat{\bm \xi}_j\|^2$ for $j \in \vx({d+1})$ inductively.

For $d\geq 1$, consider an induction hypothesis for $\vx(d)$
\begin{equation}\label{induction}
  \begin{split}
    P(\widehat \pa(k)\neq \pa^\circ(k)) &\leq  c_7 e^{-c_5n D_{\min} + \log p + n(\Delta_n-1)}, \quad 
    \|\widehat g_k - g_k^\circ \|_{L_2} =O_p(\epsilon_n),  \\
    P\left( \|\widehat{\bm \xi}_k -\bm \xi_k\|^2 \geq c_{d-1} n\epsilon_n^2 \right) 
    & \leq c_{7} e^{-c_{d-1}n\epsilon_n^2}, \quad \forall k \in \vx(d),
  \end{split}
\end{equation}
where $c_{d-1}>0$ is a constant. 

For $k\in \vx(1)$, $\bm \xi_j=\widehat{\bm \xi}_j$ and $\widehat{g}_k=g^\circ_k=0$, so the induction hypothesis \eqref{induction} is satisfied. 

For $d_j = d$, we will prove that \eqref{induction} is
met given that it is satisfied by $k \in \vx(d)$. 
Let $\delta_n^2 \geq n^{-1}\|\widehat{g}_j(\bm Y_{\vx(d)},\bm \xi_{\vx(d)}) - g^\circ(\bm Y_{\vx(d)},\bm \xi_{\vx(d)})\|^2$. 
By (4), 
\begin{equation*}
    n^{-1}\big\|\bm Y_j - \widehat g_j(\bm Y_{\vx(d)},\widehat{\bm \xi}_{\vx(d)})\big\|^2 
\leq n^{-1}\big\|\bm Y_j - g^*_j(\bm Y_{\vx(d)},\widehat{\bm \xi}_{\vx(d)})\big\|^2.
\end{equation*}
If $\delta_n^2\geq \epsilon_n^2$, then 
\begin{equation*}
  \begin{split}
     &n^{-1}\|\bm Y_j - \widehat{g}_j(\bm Y_{\vx(d)},\bm \xi_{\vx(d)})\|^2 \\
 \leq\ & n^{-1}\|\bm Y_j - g^*_j(\bm Y_{\vx(d)},\bm \xi_{\vx(d)})\|^2 
   - 2n^{-1}(\bm Y_j - \widehat{g}_j(\bm Y_{\vx(d)},\bm \xi_{\vx(d)}))^\top 
  (\bm \xi_{\vx(d)} - \widehat{\bm \xi}_{\vx(d)})\widehat{\beta}_j\\ 
     &+ 2n^{-1}(\bm Y_j - {g}^*_j(\bm Y_{\vx(d)},\bm \xi_{\vx(d)}))^\top(\bm \xi_{\vx(d)} - \widehat{\bm \xi}_{\vx(d)}){\beta}^\circ_j
 + n^{-1}\|(\bm \xi_{\vx(d)} - \widehat{\bm \xi}_{\vx(d)}){\beta}^\circ_j\|^2 \\
\leq\ & n^{-1}\|\bm Y_j - g^*_j(\bm Y_{\vx(d)},\bm \xi_{\vx(d)})\|^2
  + 3 c (\kappa^\circ)^2 \delta_n^2,
  \end{split}
\end{equation*}
where the second inequality follows from the Cauchy-Schwarz inequality. 
By Lemma \ref{lemma:variable-selection}, 
$P(\widehat{\pa}(j)\neq \pa^\circ(j))\leq c_7 e^{-c_5n D_{\min} + \log p + n(\Delta_n-1)}$
and $\|\widehat{g}_j- g_j^\circ\|^2_2\leq \delta_n^2$. 
By the triangular inequality,
\begin{equation*}
  \begin{split}
    \| \widehat{\bm \xi}_j - \bm \xi_j\| 
    &= \| \widehat{g}_j(\bm Y_{\vx(d)},\widehat{\bm \xi}_{\vx(d)}) - g^\circ_j(\bm Y_{\vx(d)},\bm \xi_{\vx(d)})\|\\
  &\leq \|\widehat{g}_j(\bm Y_{\vx(d)},\bm \xi_{\vx(d)})-\widehat{g}_j(\bm Y_{\vx(d)},\bm \xi_{\vx(d)}) \| + 
  \|\widehat{g}_j(\bm Y_{\vx(d)}, \bm \xi_{\vx(d)})-g^\circ_j(\bm Y_{\vx(d)}, \bm \xi_{\vx(d)})\|\\
  &= \| (\widehat{\bm \xi}_{\vx(d)}-\bm \xi_{\vx(d)}) \widehat{\beta}_j\| 
  + \|\widehat{g}_j(\bm Y_{\vx(d)},\bm \xi_{\vx(d)})-g^\circ_j(\bm Y_{\vx(d)},\bm \xi_{\vx(d)})\|.
  \end{split}
\end{equation*}
Note that $\|(\widehat{\bm \xi}_{\vx(d)}-\bm \xi_{\vx(d)}) \widehat{\beta}_j\|^2\leq cn \kappa^\circ \epsilon_n^2$ 
by the induction hypothesis. 
Also, by Lemma \ref{lemma:empirical-norm}, for $\delta_n^2\geq \epsilon_n^2$, we have 
$P(\|\widehat{g}_j(\bm Y_{\vx(d)},\bm \xi_{\vx(d)})-g^\circ_j(\bm Y_{\vx(d)},\bm \xi_{\vx(d)})\|^2 
\geq c n\delta_n^2)\leq c e^{ - n\delta_n^2 }$. 
Finally, let $\delta_n^2 = C \epsilon_n^2$ for a sufficiently large constant $C>0$. 
Then we have 
\begin{equation*}
    \begin{split}
        P(\widehat{\pa}(j)\neq \pa^\circ(j))\leq c_7 e^{-c_5n D_{\min} + \log p + n(\Delta_n-1)},\\
        \|\widehat g_k - g_k^\circ \|^2_2 =O_p(\epsilon_n^2),\\
        P(\|\widehat{\bm \xi}_j-\bm \xi_j\|^2\geq c_{d} n \epsilon_n^2 )\leq e^{-n c_d\epsilon_n^2},
    \end{split}
\end{equation*}
for $d_j = d$. This proves the induction hypothesis. 

 Finally, note that 
 \begin{equation*}
     P(\widehat{G}\neq G^\circ)\leq \sum_{j=1}^p P(\widehat{\pa}(j)\neq \pa^\circ(j)).
 \end{equation*}
 Then $\max_{1\leq j \leq p} \|\widehat{g}_j - g^\circ_j\|_{L_2}=O_p(\epsilon_n)$.

\textit{Step 2. Bounds for $\max_{1\leq j \leq p}\|\widehat{f}_j-f^\circ_j\|^2$.}
Suppose that $f_j$ is supported on a uniformly bounded set 
$\{\|\bm Y_{\vx(d)}\|_{\infty} \leq \rho_1\}$ for some constant $\rho_1>0$. Then,
there exists $\rho_2$ such that 
${\mathcal E}= \{ \|\bm \xi_{\vx(d)}\|_{\infty} <\rho_2 \}\supseteq \{\|\bm Y_{\vx(d)}\|_{\infty} < \rho_1\}$. 
Let $S = \{ k : \beta_{jk}^\circ \neq 0 \}\subseteq \vx(d)$. 
Note that 
\begin{equation*}
    \|\widehat{g}_j - g^\circ_j\|_{L_2}^2 \geq \int_{S^c} |\widehat g_j - g^\circ_j|^2 
dP = (\widehat{\beta}_{j,S} -\beta^\circ_{j,S})^\top \E( \I_{\mathcal E^c} \xi_{S}
\xi_{S}^\top )(\widehat{\beta}_{j,S} -\beta^\circ_{j,S}),
\end{equation*}
where
$\I_{\mathcal E^c}(\cdot)$ denotes the indicator.
Since $c_{-}\leq \lambda_{\min}(\bm\Sigma)\leq \lambda_{\max}(\bm\Sigma)\leq c_{+}$, 
this implies $\xi_S$ is not degenerated 
and $\E( \I_{\mathcal E^c} \xi_{S} \xi_{S}^\top )\geq c$ for some constant $c>0$. 
Hence, we have that 
$\|\widehat{\beta}_j-\beta^\circ_j\|^2=\|\widehat{\beta}_{j,S}-\beta^\circ_{j,S}\|^2=O_p(\epsilon_n^2)$.
If follows that $\E |\langle \xi_{\vx(d)}, \widehat{\beta}_j-\beta^\circ_j \rangle|^2=O_p(\epsilon_n^2)$.
By the triangular inequality, $\|\widehat{f}_j - f^\circ_j\|_{L_2} \leq
\|\widehat{g}_j - g^\circ_j\|_{L_2}+ \left(\E |\langle \xi_{\vx(d)}, \widehat{\beta}_j-\beta^\circ_j \rangle|^2\right)^{1/2}
=O_p(\epsilon_n)$, which completes the proof. 
\end{proof}

\begin{proof}[Proof of Theorem 4]
  The proof consists of three steps. 

\textit{Step 1. Truncation.} 
We truncate $\big\{Y_{j}^{(i)} : i=1,\ldots,n, \ j=1,\ldots,p \big\}$ to treat the unbounded issue. From (2), 
\begin{equation*}
    Y_{j}^{(i)} \mid Y_{\pa(j)}^{(i)} \sim N\Big(f_j\big(Y_{\pa(j)}^{(i)}\big),\sigma^2_j+\sigma^2_{\eta, j}\Big),
\end{equation*}
where $\sigma^2_{\eta,j}$ is the $j$-th diagonal of 
$\Sigma_\eta$. By the uniform boundedness of $f_j$, 
\begin{equation*}
    \max_{1\leq i \leq n}\max_{1\leq j \leq p}|Y_{j}^{(i)}|\leq c\sqrt{\log(np)}
\end{equation*}
almost surely for some constant $c>0$. 
Let $\widetilde{Y}_{j}^{(i)} = \sign\left(Y_{j}^{(i)}\right)\min\left(|Y_{j}^{(i)}|, B\right)$
be the truncated $Y_{j}^{(i)}$; $i=1,\ldots,n$, $j=1,\ldots,p$, where $B=c\sqrt{\log(np)}$ is a truncation constant. 
Then $\big\{\widetilde{Y}^{(i)}\big\}_{1\leq i \leq n}$ are independent and identically distributed. 
Let $P$ and $\widetilde{P}$ denote the probability for $Y_{j}^{(i)}$ and $\widetilde{Y}_{j}^{(i)}$.
Then 
\begin{equation*}
  \begin{split}
    \|f_j - f^\circ_j\|^2_{2} &\leq C\|f_j - f^\circ_j\|^2_{2} + C_1 P(\|Y_{j}^{(i)}\|_{\infty}>B) \leq  C (\|f_j - f^\circ_j\|^2_{2} + n^{-1}),
  \end{split}
\end{equation*}
where $C>0$ is a generic constant and $C_1>0$ is defined in Condition 4.
Note that $\widetilde{P}$ is supported on $[-B,B]^p$, so it suffices to consider the convergence rate 
of $\|\widehat f_j - f^\circ_j\|^2_{2}$, where 
$\widehat{f}_j$ is based on truncated data $\big\{\widetilde{Y}^{(i)}\big\}_{1\leq i\leq n}$ on a bounded domain $[-B,B]^{p}$. 

\textit{Step 2. Approximation Error.}
Note that $\mathcal F_j$ is uniformly bounded. By Theorem 1 of \citet{schmidt2019deep}, for any $0<\epsilon_n < 1/2$, 
there exists an FNN $f^*\in\mathcal{F}_j^n$ with depth $L=C_2\log(1/\epsilon_n)$, width $h=C_2\epsilon_n^{-\kappa^\circ/r}$, 
and $s=C_2\log(1/\epsilon_n)\epsilon_n^{-\kappa^\circ/r}$ such that 
\begin{equation*}
    \|f^*_j- f_j\|_{L_{\infty}([-B,B])}\leq \epsilon_n.
\end{equation*}
Then Condition 2 is satisfied. 

\textit{Step 3. Metric entropy.} 
Let $\mathcal{S}_{\infty}(u,m)$ be a $u$-cover of $\mathcal{F}_j^n$ in $\|\cdot\|_{L_{\infty}([-B,B]^p)}$. 
Define $g^+_k=g_k+u$ and $g^-_k=g_k-u$, where $g_k\in \mathcal{S}_{\infty}(u,m)$, $k=1,\ldots,m$.
Then $\{ g_1^{\pm},\ldots,g_m^\pm \}$ forms a $u$-bracket of $\mathcal{F}_j^n$. 
Hence, $H(u,\mathcal{F}_j^n)\leq H_\infty(u,\mathcal{F}_j^n)$, where $H_\infty(u,\cdot)$ denotes the entropy under the 
sup-norm. Then,
\begin{equation*}
    H_{\infty}(u,\mathcal{F}_j^n) 
\leq \dim(\theta_j) \log\left( \frac{6s}{u}(L (\frac{2s}{L-1})^{L-1} + (\frac{s(1-s^L)}{1-s})^2) \right)
\leq L h^2 \log(\frac{6s}{u}) + 4L^2h^2 \log(s).
\end{equation*}
Thus, the entropy integral in Condition 5 becomes 
\begin{equation*}
  \begin{split}
  \max_{1\leq j \leq p} \max_{\{ A_j : |A_j|\leq |\pa^\circ(j)|\}} \int^{\sqrt{2}\epsilon_n}_{\epsilon_n^2/256} H^{1/2}_{B}(u/c_1,\mathcal{B}_n(A_j) ) du
  &\leq \int^{\sqrt{2}\epsilon_n}_{\epsilon_n^2/256} h\sqrt{L} \sqrt{\log(\frac{6s}{u})}du + 8\epsilon_n Lh \log(s) \\
  &\leq C_5\epsilon_n L h \log(s)\leq C_6 \epsilon_n^2\sqrt{n}.
  \end{split}
\end{equation*} 
This implies Condition 5.

Finally, an application of Theorem 3 yields the desired result
when $L=C_2\log(1/\epsilon_n)$, $h=C_2\epsilon_n^{-\kappa^\circ/r}$, $s=C_2\log(1/\epsilon_n)\epsilon_n^{-\kappa^\circ/r}$, $\kappa_j=|\pa^\circ(j)|$, and $\|\beta_j^\circ\|_0\leq \varsigma_j \leq \varsigma^\circ$; $j=1,\ldots,p$,
which completes the proof.  
\end{proof}

\begin{proof}[Proof of Theorem 5]
  First, when $Y_j$ given $Y_{\vx(d)}$ is non-normal for $\topdep_j>d$, 
  $\vx(1),\ldots,\vx(d_{\max})$ are uniquely identifiable 
  by the same argument in the proof of Theorem 1.

  Let $d=\topdep_j$. Suppose 
  \begin{equation*}
    \E(Y_j \mid Y_{\vx(d)}) = \sum_{k\in\pa(j)} f_{j,k}(Y_k) + \langle \xi_{\vx(d)} ,\beta_j \rangle
    = \sum_{k\in\widetilde{\pa}(j)} \widetilde{f}_{j,k}(Y_k) + \langle \xi_{\vx(d)} , \widetilde{\beta}_j \rangle.
  \end{equation*} 
We will show that 
  \begin{equation}\label{equation:additive-identifiability-induction}
    [\widetilde f_{j,k}] = [{f}_{j,k}] + \sum_{j'\in \vx(d)} \gamma_{j'} [f_{j',k}],
    \quad \text{ for some } \gamma_{j'}; \quad j\in \vx(d),
  \end{equation}
by mathematical induction on $\topdep_j-\topdep_k$. 
  
  We begin with $\topdep_j - \topdep_k = 1$. 
Note that in $\langle \xi_{\vx(d)},\beta_j - \widetilde{\beta}_j \rangle$ the term containing $Y_k$
  is $\xi_k = Y_k - \E(Y_k\mid Y_{\vx(d)})$. Thus, $f_{j,k}(Y_k) - \widetilde{f}_{j,k}(Y_k) =  \gamma_{k} Y_k$,
  which implies $[f_{j,k}] = [\widetilde{f}_{j,k}]$.

  Consider $\topdep_j - \topdep_k = l > 1$. Suppose that 
\eqref{equation:additive-identifiability-induction} holds for $j',k'$ with $\topdep_{j'} 
- \topdep_{k'} < l$. Then, for the terms containing $Y_k$, we have 
  $f_{j,k} - \widetilde{f}_{j,k} 
    = \sum_{j'\in\vx(d)} \gamma_{j'} f_{j'k} - \widetilde{\gamma}_{j'} \widetilde{f}_{j'k}$. 
For $f_{j',k}\neq 0$ on the right-hand side, $\topdep_{j'}-\topdep_k < l$, so 
  \begin{equation*}
    \sum_{j'\in\vx(d)} \gamma_{j'} [f_{j'k}] - \widetilde{\gamma}_{j'} [\widetilde{f}_{j'k}] =
    \sum_{j'\in\vx(d)} \psi_{j'} [f_{j'k}],
  \end{equation*}
  for some $\psi_{j'}$; $j'\in\vx(d)$.
  Hence, $[\widetilde f_{j,k}] = [{f}_{j,k}] - \sum_{j'\in \vx(d)} \psi_{j'} [f_{j',k}]$. This leads to \eqref{equation:additive-identifiability-induction}.

  In \eqref{equation:additive-identifiability-induction}, $[\widetilde f_{j,k}]$ cannot be $[0]$ if the condition in Theorem 5 holds, 
  so $\pa(j)\subseteq \widetilde{\pa}(j)$. By symmetry, $\widetilde{\pa}(j)\subseteq \pa(j)$, which completes the proof.  
\end{proof}

\end{document}